\documentclass[reprint,twocolumn,superscriptaddress,amsmath,amssymb,aps,floatfix]{revtex4}

\usepackage{graphicx}
\usepackage{dcolumn}
\usepackage{bm}
\usepackage{miller}

\begin{document}

\preprint{APS/123-QED}

\title{Synchrotron x-ray scattering of UN and U$_2$N$_3$ epitaxial films}

\author{E. Lawrence Bright}
 \email{e.lawrencebright@bristol.ac.uk}
\affiliation{University of Bristol, HH Wills Physics Laboratory, Tyndall Avenue, Bristol, BS2 8BS, UK}
\author{R. Springell}
\affiliation{University of Bristol, HH Wills Physics Laboratory, Tyndall Avenue, Bristol, BS2 8BS, UK}
\author{D. G. Porter}
\affiliation{Diamond Light Source Ltd., Diamond House, Harwell Science and Innovation Campus, Didcot, Oxon OX11 0DE, UK}
\author{S. P. Collins}
\affiliation{Diamond Light Source Ltd., Diamond House, Harwell Science and Innovation Campus, Didcot, Oxon OX11 0DE, UK}
\author{G. H. Lander}
\affiliation{University of Bristol, HH Wills Physics Laboratory, Tyndall Avenue, Bristol, BS2 8BS, UK}

\date{\today}

\begin{abstract}
We examine the magnetic ordering of UN and of a closely related nitride, U$_2$N$_3$, by preparing thin epitaxial films and using synchrotron x-ray techniques.
The magnetic configuration and subsequent coupling to the lattice are key features of the electronic structure.
The well-known antiferromagnetic (AF) ordering of UN is confirmed, but the expected accompanying distortion at T$_N$ is not observed.
Instead, we propose that the strong magneto-elastic interaction at low temperature involves changes in the strain of the material.
These strains vary as a function of the sample form.
As a consequence, the accepted AF configuration of UN may be incorrect. 
In the case of cubic $\alpha$-U$_2$N$_3$, no single crystals have been previously prepared, and we have determined the AF ordering wave-vector.
The AF T$_N$ is close to that previously reported. 
In addition, resonant diffraction methods have identified an aspherical quadrupolar charge contribution in U$_2$N$_3$ involving the 5$f$ electrons; the first time this has been observed in an actinide compound.
\end{abstract}

\maketitle

\section{Introduction}

There is renewed interest in uranium nitride as a so-called advanced-technology fuel to replace the current standard fission fuel, UO$_2$.
This is principally due to its higher thermal conductivity, 20 W/(m-K) at 1000 K \cite{Kurosaki2000}, compared to $\sim$ 3.5 W/(m$\cdot$K) for UO$_2$ \cite{Ronchi2004}. 
In stark contrast to UO$_2$, whose thermal conductivity is entirely driven by the phononic behavior, for UN only $\sim$ 15 \% is due to any phonon contribution, and the remainder is \textit{electronic} \cite{Kurosaki2000}. 
The ability to calculate these electronic contributions and therefore make predictions about the thermal properties is complex, and attempts have been made by Yin \textit{et al.} \cite{Yin2011} and by Szpunar and Szpunar \cite{Szpunar2014}, both of which use approximations. 
In fact, the electronic structure of UN has been controversial for at least 50 years. 
Despite many studies, even the number of 5$f$ electron states, and whether they are localized or itinerant (or some mixture), is still being discussed. 
The work reported here is thus a contribution to these discussions.

We have recently succeeded in preparing thin epitaxial films \cite{LawrenceBright2018} of UN and a closely associated material cubic U$_2$N$_3$, which is almost always found in conjunction with UN, as it represents a byproduct in the oxidation process. 
We have also reported the corrosion rates (with H$_2$O$_2$) \cite{LawrenceBright2019} of these materials, and found that whereas UN is less corrosive than UO$_2$, the U$_2$N$_3$ material is at least 20 times more corrosive than UN. 
Since U$_2$N$_3$ is found at the surface of UN, this higher corrosion rate is a concern, and our work reported here suggests a possible reason for this difference.

UN has the NaCl \textit{fcc} cubic structure with $a$ = 4.89 \AA{} at 300 K. 
The susceptibility gives an effective moment ($\mu_{eff}$) of $\sim$ 2.8 $\mu_B$ and a large $\theta_p$ of $\sim$ $-$  300 K in fitting to the Curie-Weiss law. 
UN orders antiferromagnetically (AF) at T$_N$ $\sim$ 53 K with a type-I AF structure with an ordered moment ($\mu_{ord}$) of 0.75 $\mu_B$ \cite{Curry1965}. 
The large discrepancy between $\mu_{eff}$ and $\mu_{ord}$ and between T$_N$ and $\theta_p$ are not understood. 
Tro\'{c} \textit{et al.} (2016) \cite{Troc2016} have recently given an excellent summary of the properties of UN.

Much less work has been done on U$_2$N$_3$, although the structure of the cubic ($\alpha$ form) has been known for many years \cite{Rundle1948}, and is the cubic bixbyite structure common to R$_2$O$_3$, where R is a metal atom.
The lattice parameter is between 10.6 and 10.7 \AA{}, depending on the exact U/N ratio. 
Tro\'{c} (1975) \cite{Troc1975} examined the magnetic properties, and the effective moments are around 2 $\mu_B$. 
The AF ordering temperatures vary as a function of the U/N ratio between 94 K for UN$_{1.50}$ to $\sim$ 20 K for UN$_{1.72}$. 
Neither the type of AF magnetic ordering, nor the ordered moments, are known.

The preparation of such films opens the way for further experimental studies of the properties of both compounds, and thus perhaps a better understanding of the electronic structure, which can then be used in modeling for the thermal conductivity and other properties. 
In this paper we discuss experiments below room temperature on both UN and U$_2$N$_3$ epitaxial films focused on the interaction of the lattice and the magnetic (electronic) structure. 
This range of temperature is, of course, irrelevant for nuclear fuel applications, but our emphasis is on the electronic structure and how best to describe it.

\section{Sample Preparation and Experimental Procedure}

Although thin films of UN have been produced before \cite{Black2001,Rafaja2005,Zhang2010,Long2016,Wang2016,Lu2016}, they have not been epitaxial, but in the best case have been strongly textured \cite{Rafaja2005}. 
As reported in Ref. \cite{LawrenceBright2018}, we used a sapphire \hkl(1-102) substrate with a \hkl[001]-oriented Nb buffer, and the UN grows on this with a 1:$\sqrt{2}$ relation and a 45 $^\circ$ rotation. 
The growth temperature of the film was 600 $^\circ$C, and a 5 nm cap of Nb was deposited at room temperature on the film to prevent oxidation. 
The film used at the synchrotron had a thickness of 70 nm and a rocking curve of 1.9 $^\circ$. 
Thin films of U$_2$N$_3$ have not been reported previously, and we found these can be grown on CaF$_2$ substrates ($a_0$ = 5.451 \AA{}) and have a very good mosaic (less than 0.10 $^\circ$) \cite{LawrenceBright2018} when they are thin. 
However, for thicker films the mosaic increases, and the 200 nm film we used had a rocking curve of $\sim$ 1 $^\circ$. 
The lattice parameter in the direction of growth is 10.80(1) \AA{}, compared to $2a_0$ (CaF$_2$) = 10.9 \AA{}, and the in-plane parameters were 10.60(2) \AA{}. 
Based on the atomic volume, this corresponds to an effective cubic lattice parameter of 10.67 \AA{}, suggesting we are close to stoichiometry \cite{Troc1975}. 

Resonant x-ray scattering (RXS) measurements were conducted at the Materials and Magnetism Beamline I16 at Diamond Light Source \cite{Collins2010}. 
The x-ray energy was tuned to 15 keV ($\lambda$ = 0.8265 \AA{}) for sample alignment and determination of the lattice parameter, due to the increased number of reflections available, and to the uranium $M_4$ edge at 3.726 keV ($\lambda$ = 3.327 \AA{}) for measurements of the magnetic diffraction, taking advantage of the resonant enhancement of the magnetic signal. 
Samples were mounted in a closed-cycle cryocooler for low temperature measurements. 
A kappa-geometry 6-circle diffractometer provides large access to reciprocal space and the capability of azimuthal scans and grazing incidence diffraction. 
All measurements were performed in vertical geometry, perpendicular to the incident polarization of the beam and the azimuthal zero reference is taken when the crystal \hkl(001) direction intersects the scattering plane. 
Scattered x-rays were measured using either the high-sensitivity Pilatus3-100K photon-counting area detector or by scattering at $\sim$ 90 $^\circ$ from a graphite analyzer crystal into a photo-diode. 
Rotating the analyzer crystal about the scattered wave-vector provides a measurement of the polarization of the diffracted signal.

\section{Results and Discussion}

\subsection{Structural properties of UN films}

\begin{figure}[htb]
\includegraphics[width=1\linewidth]{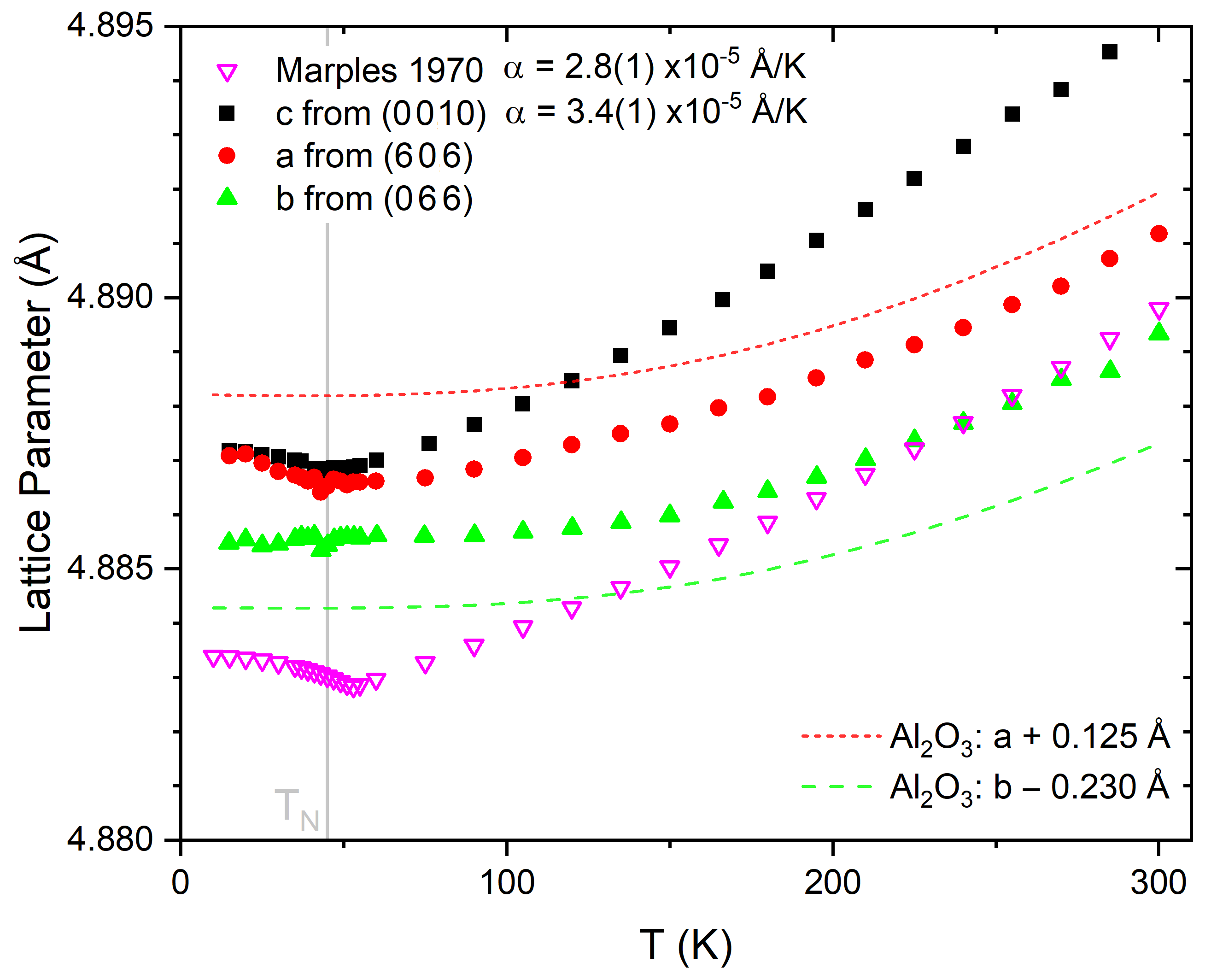}
\caption{\label{fig:unlat} Lattice parameters as measured from different reflections for the UN film. The values reported by Marples 1970 [18] are given by inverted triangles. $c$ is defined as the lattice parameter in the growth direction of the film. The UN film thickness is 200 nm and the film is deposited on a Nb \hkl(001) buffer on a sapphire \hkl(1-102) substrate \cite{LawrenceBright2018}.}
\end{figure}

Figure \ref{fig:unlat} shows the variation of the lattice parameters in the UN film as a function of temperature measuring higher-order Bragg reflections. 
The values given by Marples (1970) \cite{Marples1970}, measured from a polycrystalline sample, are shown as inverted triangles. 
From this it can be seen that our UN films are slightly \textit{expanded} by $\sim$ 0.004 \AA{} ($\sim$7 $\times$ 10$^{-4}$ in terms of strain) in the growth direction due to the interaction with the buffer and substrate. 
The lattice linear expansion coefficient from 100 - 300 K is approximately the same as that reported in Ref. \cite{Marples1970}, but from the in-plane lattice components we can see that the film is under tensile strain of $\sim$ +20 $\times$ 10$^{-6}$, where the growth axis is larger than the mean of the in-plane parameters. 
Moreover, this strain increases with temperature, as the expansion of the sapphire substrate (indicated by dashed lines in the figure) is smaller than that of UN. 

A further point to make here is to note the \textit{expansion} of the lattice below the AF ordering temperature (T$_N$). This will be considered more carefully below, but it represents an important measure of the interaction of the UN lattice with the magnetic (electronic) components. 

\begin{figure}[htb]
\includegraphics[width=1\linewidth]{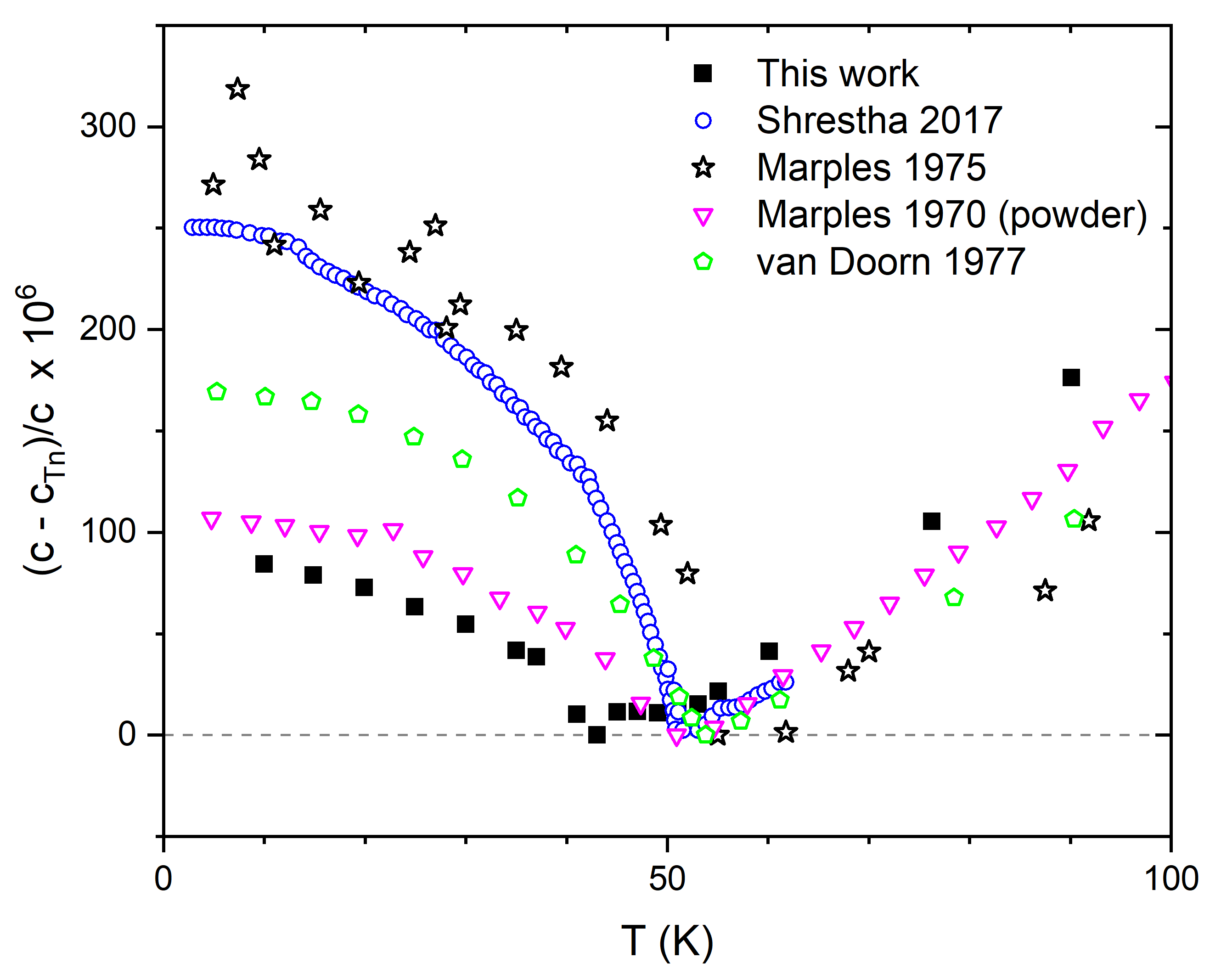}
\caption{\label{fig:unlitcom} Relative variation of the lattice parameters in UN below T$_N$ as measured in our work and previously reported in Refs. \cite{Marples1970,Marples1975,Doorn1977,Shrestha2017}.}
\end{figure}

Figure \ref{fig:unlitcom} focuses on the expansion in the UN lattice when the material orders. 
We find that our film has T$_N$ = 45.8(3) K, which is lower than the 52-55 K region found in bulk samples \cite{Troc2016}, and is not surprising given the effect of the strain induced by the substrate. 
This figure includes data from previous studies \cite{Marples1970,Marples1975,Doorn1977,Shrestha2017}. 
Apart from Marples  \cite{Marples1970}, all studies were performed on single crystals, although Refs.  \cite{Doorn1977,Shrestha2017} used strain-gauge techniques, not x-rays. 
What is surprising about this figure is that the expansion of the lattice appears to depend on the sample form, and the magnitude is thus \textit{not} a true bulk property, as it varies by almost a factor of three between different samples. 
Our results give a lower value, similar to that derived from a polycrystalline sample as measured by Marples  \cite{Marples1970}. 
This is particularly interesting, as the study by Shrestha \textit{et al.}  \cite{Shrestha2017} shows that this expansion may be partially suppressed by the application of a modest magnetic field (< 30 T), although there seems no obvious explanation for this in the AF state. 
Magnetic fields of $\sim$ 60 T are required  \cite{Troc2016,Shrestha2017} to disrupt the AF order of UN.

\textit{We now come to the question of a lattice distortion at T$_N$.} 
Curry was the first to report the AF structure of UN in 1965  \cite{Curry1965}; the structure consists of ferromagnetic sheets of uranium moments arranged in a simple $+$ $-$ orientation along the propagating axis, which, in the single-\textbf{k} form, may be any one of the cube axes \hkl<100>. 
The moments are parallel to the propagation direction. 
This immediately gives three possible domains (neglecting time reversal), and the symmetry is \textit{tetragonal}, i.e. one would expect a distortion at T$_N$ so that $c$ (parallel to the propagation direction) is no longer equal to $a$ and $b$ (perpendicular to the propagation direction). 
However, the possibility of a so-called 3\textbf{k} structure, in which all domains exist in one unit cell, cannot be distinguished by the intensities of the reflections, \textit{and this 3\textbf{k} configuration has cubic symmetry}. 
Rossat-Mignod \textit{et al.}  \cite{Rossat-Mignod1980} were the first to test this on UN and concluded that UN was indeed a 1\textbf{k} system, at least under the application of uniaxial stress. 
They stated that with uniaxial stress UN became tetragonal with $c/a$ \textgreater 1. 
Marples \textit{et al.}  \cite{Marples1975} looked specifically with x-ray diffraction for the expected distortion and reported that $c/a$ = 0.99935(3) at the lowest temperature, i.e. the resulting strain $2(c - a)/(c +a)$ = – 6.5 $\times$ 10$^{-4}$. 
Note this is the \textit{opposite} sign to that suggested in Ref.  \cite{Rossat-Mignod1980}. 
A distortion implies that different $d$-spaces will be detected; for example, in the case of the \hkl(0 0 10) reflection that the $d$-space for \hkl(0 0 10) will be different from that for \hkl(10 0 0) and \hkl(0 10 0) reflections. 
A subsequent study, also on a single crystal, by Knott \textit{et al.} \cite{Knott1980}, found a smaller broadening of the full-width at half maximum (FWHM) than reported in Ref. \cite{Marples1975} and concluded that the distortion, if present, was smaller than reported in  \cite{Marples1975}. 
There is, therefore, some doubt over the existence of such a tetragonal distortion.

\begin{figure}[htb]
\includegraphics[width=1\linewidth]{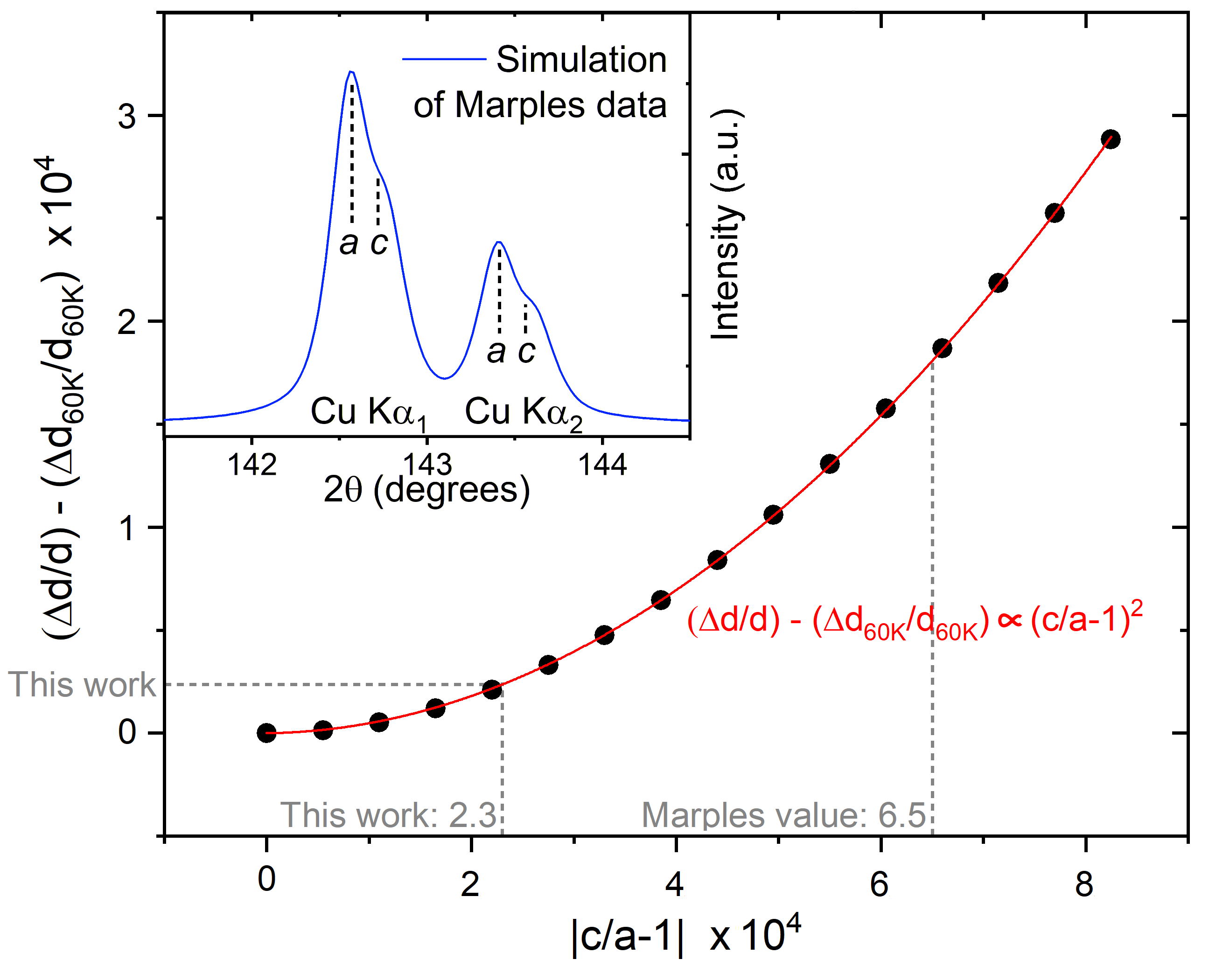}
\caption{\label{fig:unsim} The relationship between the tetragonal distortion, $\left|c/a - 1\right|$, versus the relative change of the FWHM ($\Delta d/d$) obtained in a simulation for UN. The proportionality factor is 417. The insert shows the profiles that would have been measured by Marples \textit{et al.} (1975)  \cite{Marples1975} if ($c/a - 1$) = – 6.5 $\times$ 10$^{-4}$. Our data shows that any distortion is smaller, $\leq$ 2.3 $\times$ 10$^{-4}$ for $\left|c/a - 1\right|$.}
\end{figure}

Synchrotron x-rays have the advantage over laboratory source x-rays that there is only one single wavelength and not a mixture (e.g. Cu K$\alpha_1$ and Cu K$\alpha_2$) in the beam, so we have used this to lower the limit found by samples to a possible strain of $\sim$ 2 $\times$ 10$^{-4}$, as shown in Figure \ref{fig:unsim}. 
Unfortunately, Marples \textit{et al.}  \cite{Marples1975} do not show their raw data of the diffraction profiles, which are simulated in our figure. 
However, they do show the broadening of the FWHM of their peaks, which they then analyze in terms of a distortion. 

\begin{figure}[htb]
\includegraphics[width=1\linewidth]{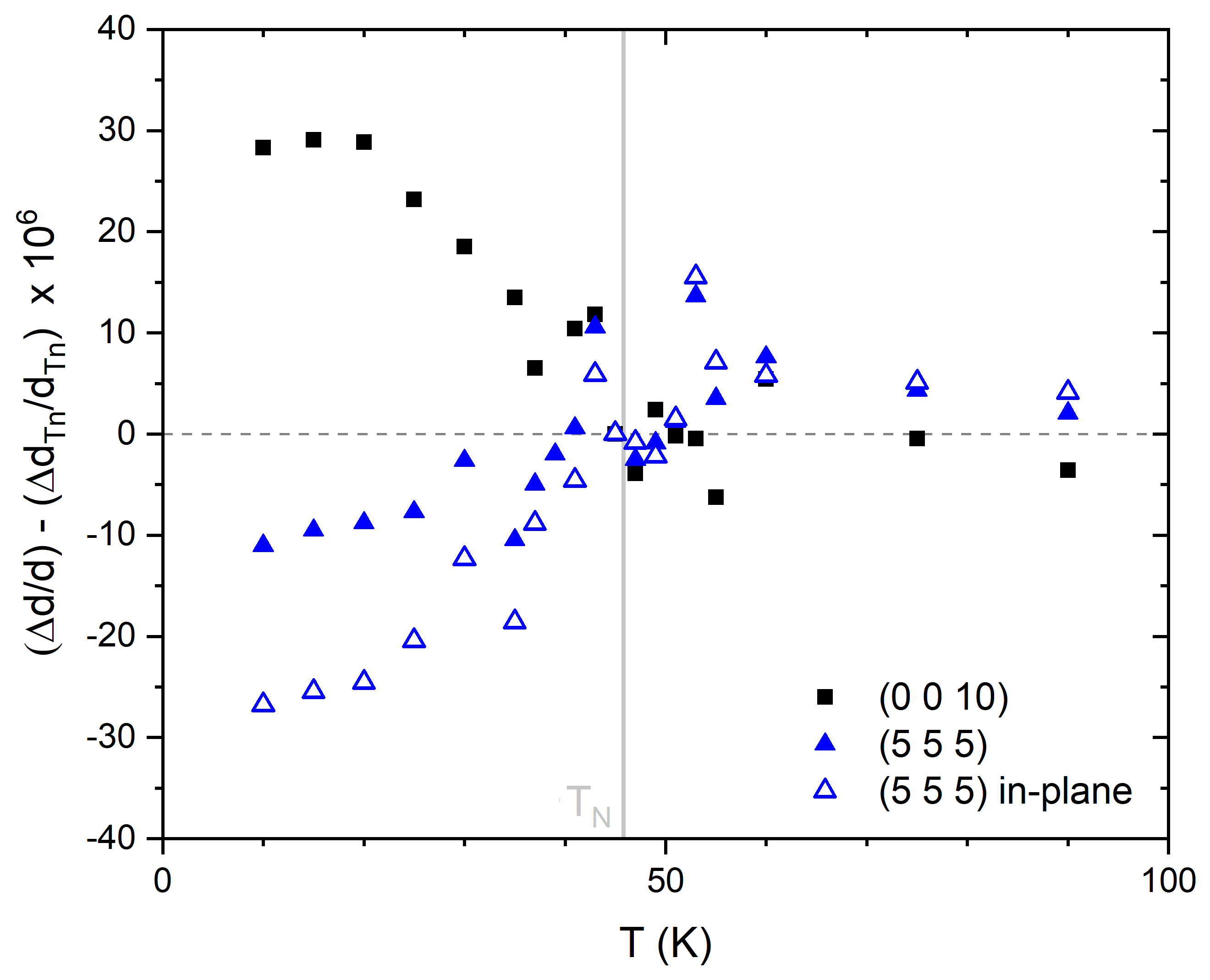}
\caption{\label{fig:unfwhm} The relative strain below T$_N$ for the UN film sample in two different directions. The growth direction is given by the \hkl(0 0 10) reflection, showing a tensile (expanding) strain as the temperature is lowered below T$_N$. The in-plane strain (which is compressive in nature) is deduced from combining the results from the \hkl(0 0 10) and \hkl(555) reflections, and is reduced below T$_N$. The error bars are $\pm$ 5 $\times$ 10$^{–6}$ units.}
\end{figure}

However, broadening of the peaks can also be a result of changing strain. 
This is shown dramatically in Figure \ref{fig:unfwhm}, where we show what happens to the FWHM of the \hkl(0 0 10) and \hkl(555) reflections from the film. 
The strain along the $c$ axis (growth direction) of the film increases, whereas the corresponding in-plane strain actually compensates by \textit{decreasing}. 
Clearly, this is a complicated effect, driven by the film-substrate interaction, but it gives no support to the idea of a tetragonal distortion in UN. 
If that were the case the \hkl(555) reflection should not change its FWHM, as all $d$-spaces for this reflection are the same whether a tetragonal distortion occurs or not. 
Thus, the \hkl(555) reflection could change its position, but should not broaden; instead it actually narrows its FWHM with decreasing temperature below T$_N$.

The combination of Figs. 2 - 4 suggests that the tetragonal distortion in UN may not be present without the external perturbation of uniaxial stress, and that strain effects in the different samples are more important. 
It is possible, therefore, that the true state of the magnetic configuration is 3\textbf{k}, where a tetragonal distortion would not be expected. 
The only \textit{unique} way to distinguish these two possibilities is by analysis of the polarization of the spin waves, a complicated neutron inelastic experiment performed only so far for UO$_2$  \cite{Blackburn2005} and USb  \cite{Magnani2010}, both 3\textbf{k} systems.

\subsection{Magnetic Scattering from UN and U$_2$N$_3$ films}

\subsubsection{UN}

One of the possibilities with the UN epitaxial film was that the strain in the lattice because of the interaction with the substrate might induce a \textit{single magnetic domain to be observed}. 
However, below T$_N$ eleven different magnetic reflections (all related by the known magnetic wave-vector of \textbf{q} = \hkl<001>) were observed when the energy was changed to the U $M_4$ edge, and no absences were found. 
For example in a 1\textbf{k} configuration, the \hkl(110) belongs to a $c$ domain (corresponding to the propagation direction) along \hkl[001], whereas the \hkl(101) reflection corresponds to a $b$ domain, and the \hkl(011) to the $a$ domain. 
All were present. 
Thus the hope that a change in the population of the 1\textbf{k} domains might be induced by the substrate-film interaction inducing a slightly “orthorhombic” UN (as noted in Fig. \ref{fig:unlat}) was not fulfilled. 
On the other hand, if the true configuration is 3\textbf{k}, all such reflections would be present. 
However, this observation is not \textit{proof} of a 3\textbf{k} AF state.

The exact intensity in resonant x-scattering (RXS) is complicated, and not related directly to the value of the magnetic moment \cite{Hill1996}. 
The observed intensities depend greatly on the large absorption, which at the U $M_4$ energy (3.726 keV) in UO$_2$ can reach values of $\sim$ 5 $\times$ 10$^4$ cm$^{-1}$  \cite{Cross1998} corresponding to $f''$ (the imaginary part of the structure factor) reaching $\sim$ 70 electrons. 
In UN this corresponds to a 1/e attenuation length of $\sim$ 150 nm. 
Some of the beam will pass through the 70 nm film of UN at this energy, but the absorption will depend on the precise geometry and is a difficult correction to make. 
It is noteworthy that (so far) no report relating \textit{intensities} of magnetic reflections measured at the U $M_4$ edge has been published. 
These arguments apply also to U$_2$N$_3$, and will not allow the magnetic structure to be determined in that material with this RXS technique. 
Such an investigation with RXS was reported by Watson \textit{et al.} (1998) \cite{Watson1998}, but at the $L_3$ edge of Nd, where the energy is higher than that at the U $M_4$ edge, and the resonant absorption (i.e. $f''$) much smaller.

\begin{figure}[htb]
\includegraphics[width=1\linewidth]{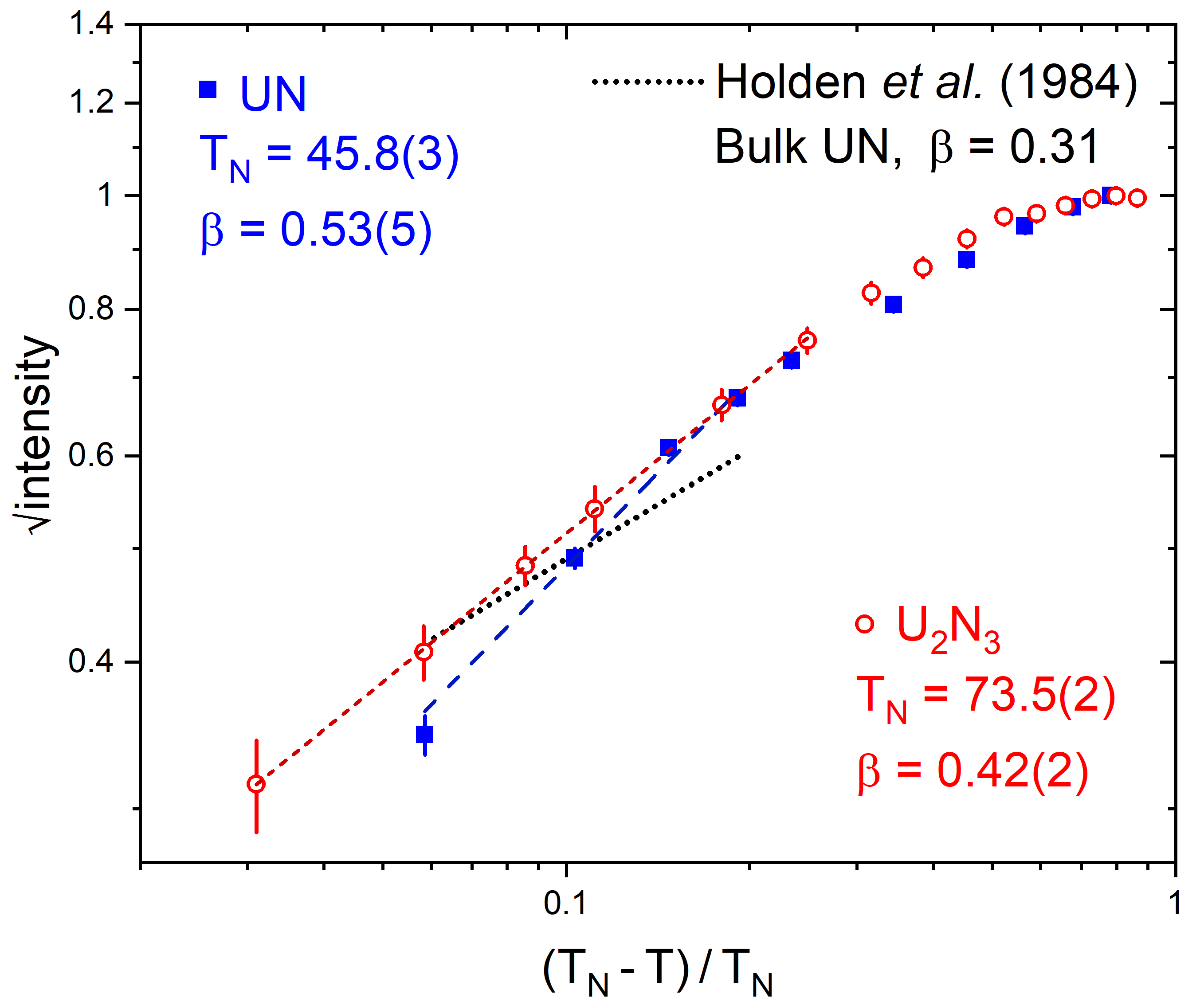}
\caption{\label{fig:beta} Plot of the square-root of the integrated intensity of the \hkl(001) for UN and \hkl(003) for U$_2$N$_3$ as a function of reduced temperature t = (T$_N - $T)/T$_N$ on a log-log plot to determine T$_N$ and $\beta$. The dotted black line gives $\beta$ = 0.31 as determined for bulk UN Ref. \cite{Holden1982}.}
\end{figure}

Figure \ref{fig:beta} shows the variation of the intensities of magnetic reflections from the two materials as a function of temperature. 
The reflections are \hkl(001) for UN and \hkl(003) for U$_2$N$_3$. 
Previous work on bulk UN \cite{Holden1982} has given a value of $\beta$ = 0.31(3), and early work on a similar bulk rocksalt uranium compound USb \cite{Lander1978} gave a value of 0.32(2). 
The value determined here, 0.53(5), for UN appears significantly higher. 
However, there is evidence that critical exponents from thin films are not necessarily the same as those determined from bulk samples. 
A good example is our recent work on UO$_2$ \cite{Bao2013}, where the values for thin films range considerably in value, and indicate a 2$^{nd}$-order phase transition, whereas bulk UO$_2$ has a 1$^{st}$-order transition at T$_N$. 
The value found for U$_2$N$_3$ is consistent with a simple mean-field model for the transition.

\begin{figure}[htb]
\includegraphics[width=1\linewidth]{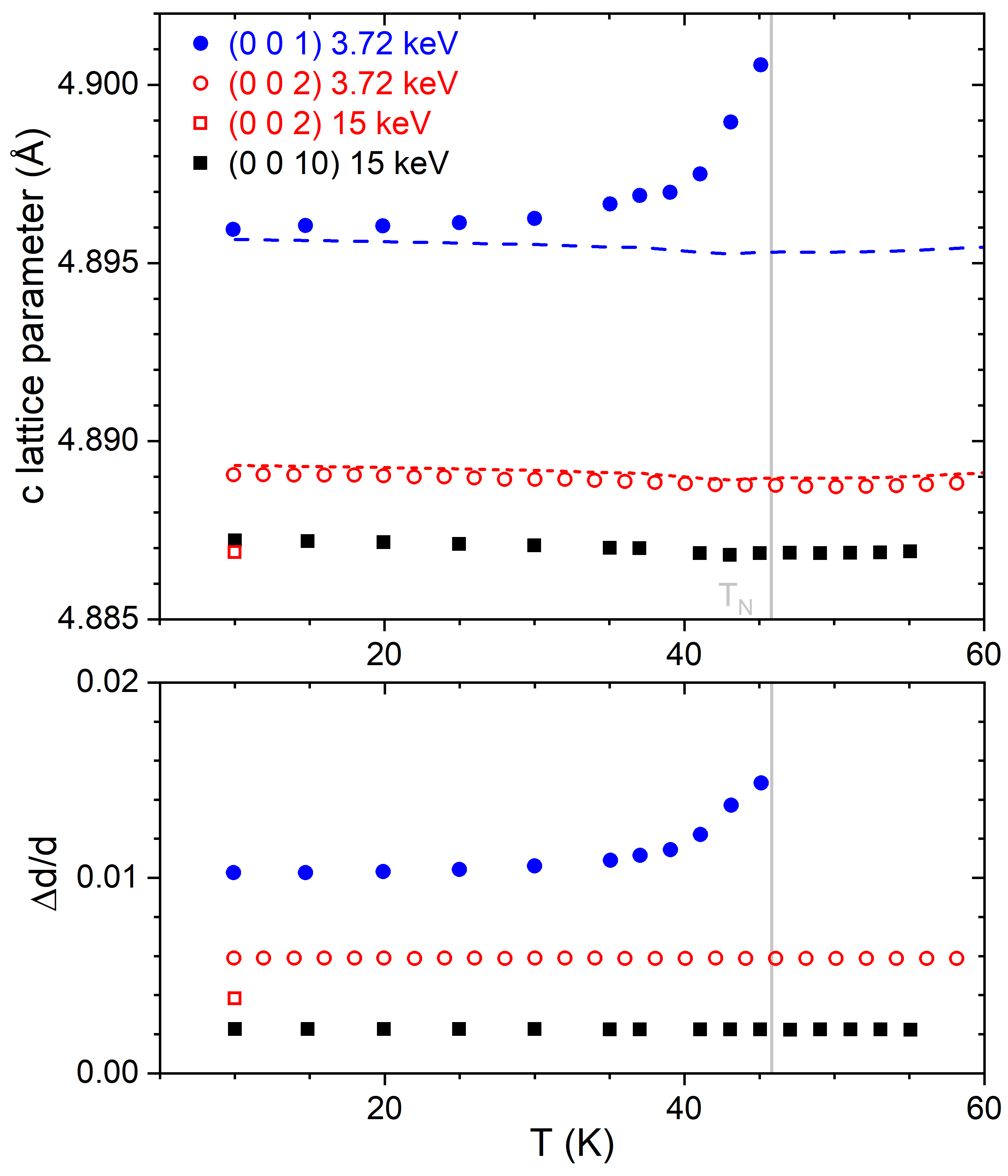}
\caption{\label{fig:unrefrac} The lattice parameter $c$ and the FWHM ($\Delta d/d$) observed for the various specular peaks as a function of temperature. The large differences in measured $c$ lattice parameter, when the energy is shifted from 15 keV to the resonant energy of 3.72 keV, is caused by refraction. The expected values of the \hkl(001) and \hkl(002) at 3.72 keV, assuming the values at 15 keV are correct, are given as dashed lines using $\delta$ = 2.0 $\times$ 10$^{-4}$, see text.}
\end{figure}

Figure \ref{fig:unrefrac} shows the lattice parameters extracted from longitudinal scans of the specular reflections, as a function of temperature, together with their \textit{relative} widths $\Delta d/d$. 
All the reflections taken at 15 keV are in good agreement with one another (as they should be), but lattice parameters measured at the resonant energy appear to be greater. 
This is due to \textit{refraction effects} and has been known for many years \cite{James1965}. 
Normally, these effects are small, but we have an \textit{unusual} case of using relatively long wavelengths x-rays, $\lambda$ = 3.327 \AA{}, at the U $M_4$ resonant edge, and the electron density per unit cell, $n_e$ = 3.41 electrons/\AA{}$^3$, is high because of the uranium. 

For specular type reflections Greenberg (1989) \cite{Greenberg1989} has shown that Bragg’s law can be re-written as
\begin{equation}
\lambda = 2 d \,  sin \theta (1 - \delta / sin^2 \theta),
\end{equation}
where $\delta$ is the correction to the refractive index defined as $n = 1 - \delta$, assuming $n = 1$ in vacuum. 
This may be readily transformed in the simple case of a specular reflection from sets of planes perpendicular to the growth direction to note that the change in the effective $c$ lattice parameter $\Delta c$ is given by
\begin{equation}
    \Delta c / c = \delta / sin^2 \theta'.
\end{equation}
$\delta = r_0 \lambda^2 n_e / (2 \pi)$, where $r_0$ is the classical electron radius, $r_0$ = 2.82 $\times$ 10$^{-5}$ \AA{}, giving a value of $\delta$ = 1.7 $\times$ 10$^{-4}$. 
Normally these values seldom exceed a few parts in 10$^{-5}$. 
Here we have also changed $\theta$ to $\theta'$ to reflect the fact that we are very close to, but not actually at, the specular condition. 
This is because of the miscut in the substrate. 
$\theta'$ represents the angle between the beam and the surface of the film.

Such effects are much more noticeable for low-angle reflections as the effect is proportional to $1/sin^2 \theta$, which implies [given the almost linear relationship for the low-angle reflections between the reflections \hkl(00L) and $sin \theta$ for the Bragg reflections] that the effect for the \hkl(001) reflection is $\sim$ 4 times more pronounced than for the \hkl(002). 
The dashed lines in Fig. \ref{fig:unrefrac} show the values derived by assuming that the 15 keV data give the true value, and the refractive index correction has a value of $\delta$ = 2.0 $\times$ 10$^{-4}$. 
Given the approximations made with the miscut and treatment of absorption at the lower energy, the agreement with calculated value (1.7 $\times$ 10$^{-4}$) is satisfactory.

By comparison for the \hkl(0 0 10) reflection using 15 keV x-rays, the correction $\Delta c/c$ $\sim$ 0.15 $\times$ 10$^{-4}$, which is smaller than the error bars in Fig. \ref{fig:unrefrac}.

Finally, we note in Fig. \ref{fig:unrefrac} an unusual effect for the \textit{magnetic} \hkl(001) reflection as a function of temperature. 
There appears to be a steady increase in the effective $c$ lattice parameter around T$_N$, and this is accompanied by a systematic \textit{increase} in the width of the reflections $-$ signifying a decreased correlation length in the critical regime of UN as the sample is warmed through T$_N$. 
This shift cannot, of course, be due to refraction, as the wavelength does not change in these measurements. 
A simple explanation might be that the magnetic correlations become incommensurate with the underlying lattice, but in that case \textbf{two} diffraction peaks would be observed. 
The \hkl(001) magnetic peak in UN arises from the reciprocal lattice points \hkl(000) $+ q_m$ and \hkl(002) $- q_m$, where $q_m$ is the magnetic wave-vector, and if $\left|q_m\right| \neq$ 1, then two peaks would be observed, symmetric about the \hkl(001) position. 
There is no sign of two such peaks. 
Instead we have a small shift (Dq) in the parameter coupling the magnetic Bragg peak to the lattice; it is as if the magnetic correlations are connected to a lattice with a slightly different (larger) spacing.

This unusual effect has been observed previously at the U $M_4$ edge, Bernhoeft \textit{et al.} (2004) \cite{Bernhoeft2004} with the compound USb. 
We shall not discuss this at length here, as Ref. \cite{Bernhoeft2004} gives a general overview of experiments on various samples, and proposes an explanation, albeit complicated, to understand this shift. 
Subsequent to this work in 2004, an effort was made to see whether the shifts could also be observed with neutron diffraction, where the resolution is not normally as good as with synchrotron x-rays. 
The successful observation with neutrons, Prokes \textit{et al.} (2009) \cite{Prokes2009}, demonstrates that the effect is not related to the surface of the sample, nor is it a property unique to actinide compounds. 
We note that this effect is always associated with an \textit{increase} in the width of the magnetic diffraction peaks, signifying a reduced correlation length. 
This may be seen clearly by noting that the upturn in both panels of the \hkl(001) position and relative width in Fig. \ref{fig:unrefrac} occur near T$_N$.

\begin{figure}[htb]
\centering
\includegraphics[width=0.9\linewidth]{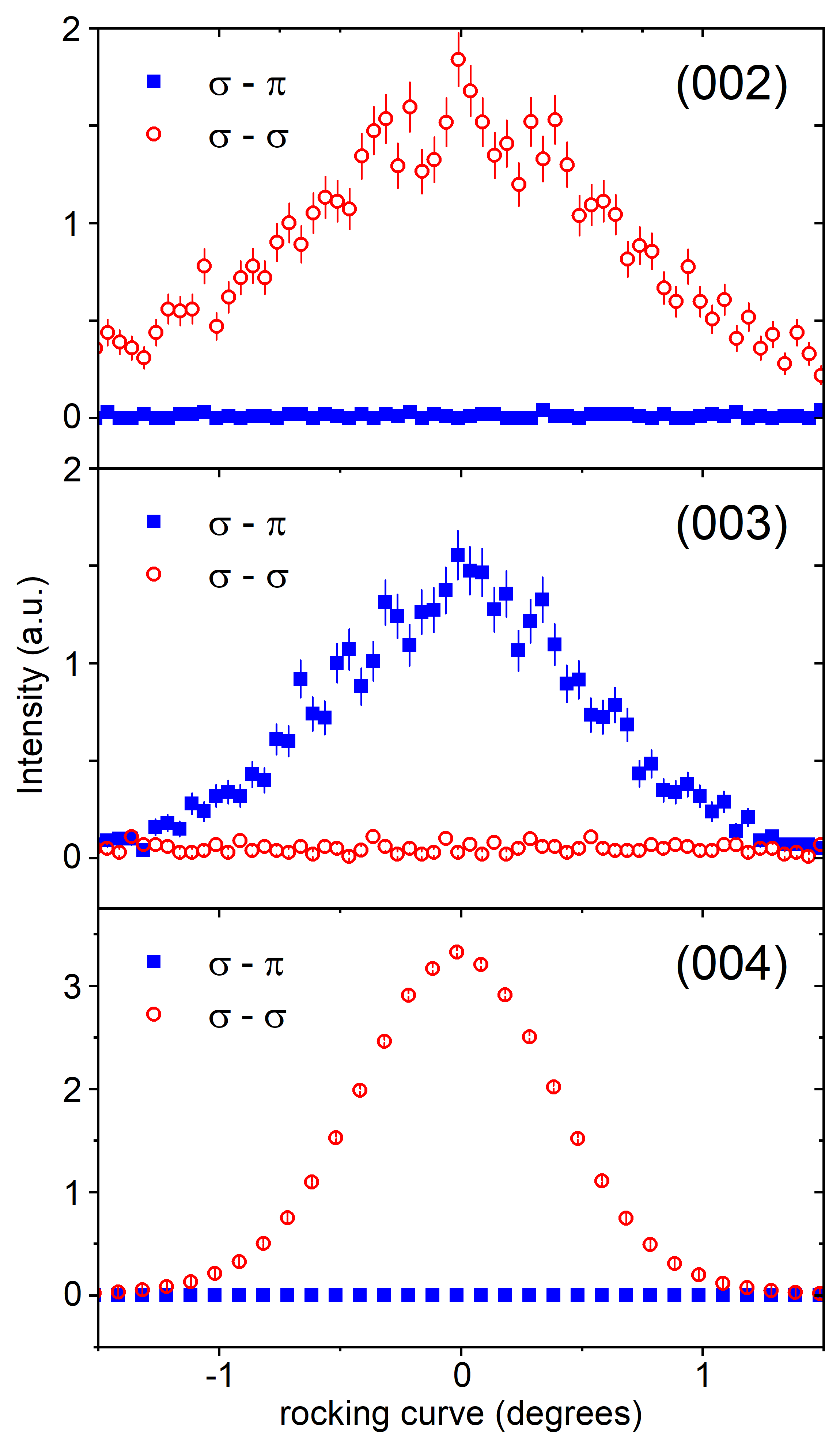}
\caption{\label{fig:u2n3pol} Polarization dependence of the specular peaks \hkl(00L) in U$_2$N$_3$ at 10 K measured at the U $M_4$ resonance energy. The data shows a complete separation between the charge peaks from the \textit{bcc} structure and the magnetic peaks arising from the ordered moments below T$_N$ = 73 K. The \hkl(002) is a weak reflection; its presence signifies that the positional parameter for the U$_2$ atom is different from zero. The \hkl(003) reflection is magnetic and disappears at T$_N$.}
\end{figure}

\subsubsection{U$_2$N$_3$}

The magnetic structure of U$_2$N$_3$ has magnetic peaks that appear at 73.5 K (see Fig. \ref{fig:beta}) in positions in which $h$ + $k$ + $l$ = odd, i.e. they do not overlap with the charge reflections from the \textit{bcc} structure where the $h$ + $k$ + $l$ = even. 
The magnetic wave vector is \textbf{q} = \hkl<001>. 
This implies that the uranium atoms related by the \textit{bcc} translation have \textit{oppositely} directed magnetic moments. 
Figure \ref{fig:u2n3pol} shows the polarization analysis scans of three different reflections, one magnetic, and two charge, along the specular direction \hkl[001] of the U$_2$N$_3$ film at 10 K. 
Purely magnetic scattering in this configuration is $\sigma$ to $\pi$, and purely charge scattering is $\sigma$ to $\sigma$.

The bixbyite cubic structure of U$_2$N$_3$ also exists with transition metal ions, i.e. $\alpha$-Mn$_2$O$_3$.
However, these materials often have a crystallographic distortions associated with the ordering, \cite{Cockayne2013} for example. 
Since we have not detected any distortion of the unit cell [note the symmetric shape of the \hkl(004) reflection in Fig. \ref{fig:u2n3pol}], it may be more appropriate to consider the trivalent rare-earth systems, e.g. Er$_2$O$_3$ and Yb$_2$O$_3$, as investigated by Moon \textit{et al.} (1968) \cite{Moon1968}.

The magnetic structure of U$_2$N$_3$ is similar to that found in Yb$_2$O$_3$ \cite{Moon1968}; reference to Fig. 3 of that paper shows a complex non-collinear magnetic structure with the moments directed along their local symmetry axes. 
Of course, since Yb$_2$O$_3$ orders at 2.25 K, the exchange interactions in U$_2$N$_3$ and Yb$_2$O$_3$ are clearly different (For a start, Yb$_2$O$_3$ is an insulator, U$_2$N$_3$ is a semi-metal), so there is no reason to expect similar magnetic structures. 
Normally, ordering temperatures in the actinides are higher than those of isostructural compounds in the rare-earths, simply because of the larger spatial extent of the 5$f$ electrons in the actinides, and the fact that such electrons often lie close to E$_F$, whereas the 4$f$ electrons of the rare-earths lie well below E$_F$.

\subsection{Energy dependence of scattering from UN and U$_2$N$_3$ films}

\begin{figure}[htbp]
\includegraphics[width=1\linewidth]{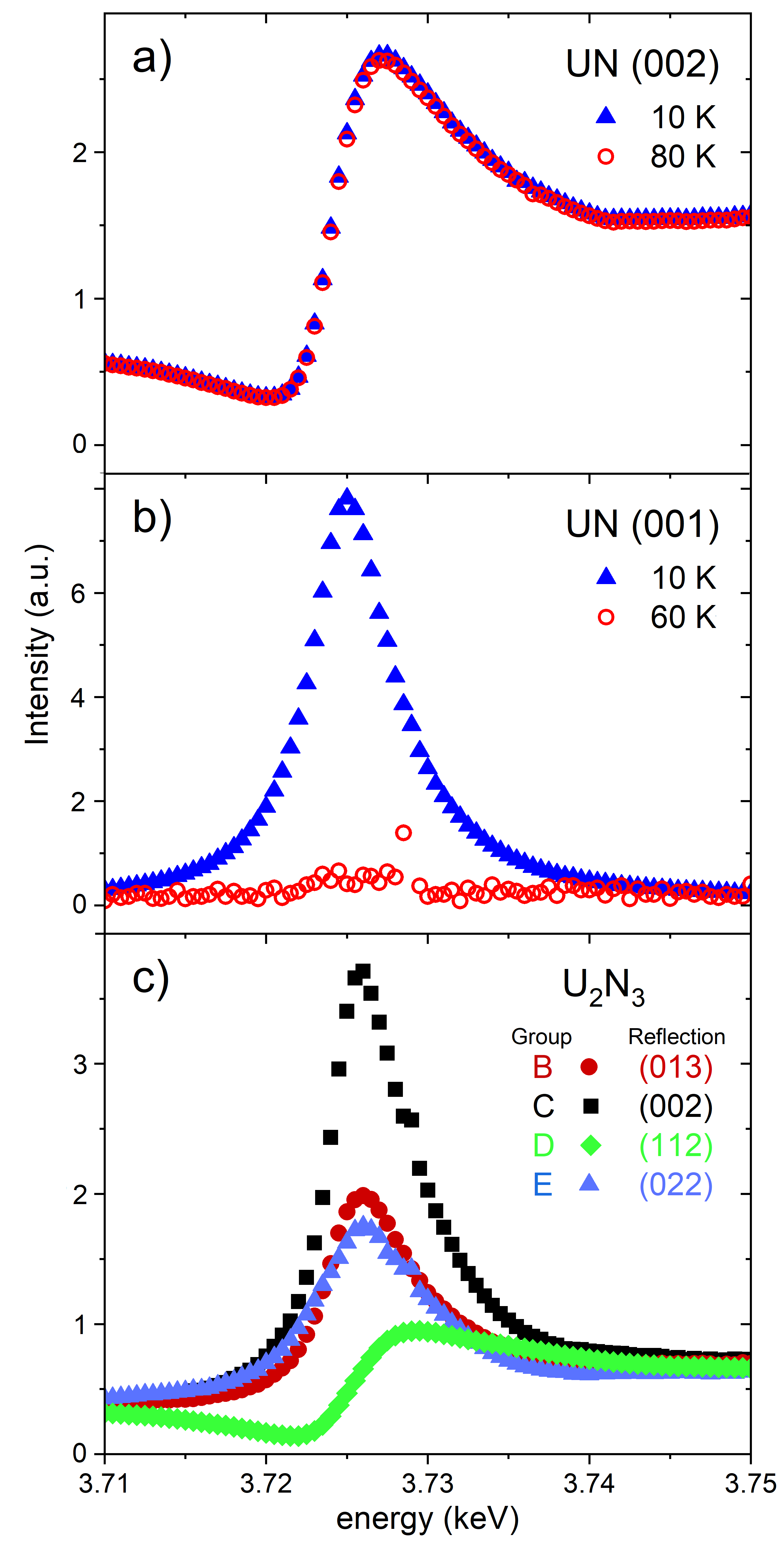}
\caption{\label{fig:en} Energy dependence of the diffraction peaks in both UN and U$_2$N$_3$. (a) and (b) show that there is no signal at the \hkl(001) of UN at T = 80 K (T \textgreater T$_N$), and that the energy dependence of the charge \hkl(002) does not change with temperature. The bottom panel (c) shows how charge peaks from U$_2$N$_3$ vary in energy according to the groups listed in Table \ref{table:sf}. Groups B, C, and E in U$_2$N$_3$ sense the $f''$ term of the scattering factor, the same term observed in magnetic scattering. Reflections in Groups A (not shown) and D have a standard charge profile in energy through the resonance. These profiles are independent of temperature.}
\end{figure}

As is well known \cite{Hill1996}, the magnetic scattering from the E1 dipole term, corresponding to the $M$ edges of uranium and dipole transitions between the filled 3$d$ core level and the partially filled U 5$f$ shell, can be represented by a complex quantity, and thus couples to the $f''$ term of the scattering factor. 
The imaginary part (related directly to the absorption) is large at the absorption edge. 
We show in Figure \ref{fig:en} (top 2 panels) the energy dependence for a charge \hkl(002) and magnetic \hkl(001) reflection in UN at base temperature. 
The \hkl(002) reflection has a standard charge profile (corresponding to the real part $f_0$ + $f'$), whereas the \textit{magnetic} \hkl(001) reflection has an energy dependence corresponding to the $f''$ term. 
The highly symmetric curve with a FWHM $\sim$ 6 eV is typical for work with thin films \cite{Bernhoeft1998}, and also reflects the partial coherence of synchrotron beam at this energy. 
Previous experiments on I16 \cite{Bao2013} have shown that the energy width can be used to determine whether the film is ordered throughout its depth. 
In this case the $\sim$ 6 eV FWHM is expected, so the film is fully ordered.

A key question the experiments above have not answered is whether both uranium sites in the cubic U$_2$N$_3$ structure order magnetically. 
It would be possible to answer this if we could reliably make the absorption corrections, but this is not the case with a 200 nm film and an absorption attenuation length of about the same order of magnitude, as discussed above. 

It is instructive at this stage to consider the geometric structure factors governing the \textit{charge} peaks in U$_2$N$_3$, and these are shown in Table \ref{table:sf} for the first 11 reflections arranged in order of $Q$, the momentum transfer. 
In comparing with experiments note that the system is cubic so \hkl(hkl) values can be permuted.

There are two uranium sites in U$_2$N$_3$, which adopts the structure of the centro-symmetric space group \# 206. 
Eight U$_1$ atoms in the unit cell sit on sites with three-fold rotational inversion symmetry (C$_{3i}$), and twenty-four U$_2$ atoms sit on sites with two-fold rotational symmetry (C$_2$). 
There is one adjustable positional parameter for the U$_2$ sites with positions \hkl(x 0 \frac{1}{4}) etc., and none for the U$_1$ atoms with positions \hkl(\frac{1}{4} \frac{1}{4} \frac{1}{4}) etc. 
X-ray \cite{Masaki1972} and neutron diffraction \cite{Tobisch1967} have been used to determine the positional parameter for U$_2$ and the consensus value is $x = -0.02$, which we have used in the calculations below. 
The presence of a finite charge intensity (i.e. $\sigma$ to $\sigma$ scattering) at the position \hkl(002) [Fig. \ref{fig:u2n3pol}] is direct proof that $x$ for U$_2$ is not zero.

This table is simply the geometric term in the structure factor \textit{only for the uranium ions}. 
The 48 N atoms will, of course, contribute to the total intensity, but only by a small amount, and we can neglect this.

Our initial interest in these reflections was in those of group E, where the contributions from the two different U atoms cancel. 
Of course, this statement is true for the spherical charge density contributions, but aspherical electron distributions, arising from quadrupoles, will stand out after the spherical part cancels. 
The symmetry elements of both U sites in U$_2$N$_3$ allow an interesting phenomenon called anisotropic tensor scattering (ATS) \cite{Collins2001,Kokubun2012} to be observed. 
The symmetry implies that aspherical (quadrupolar) electron distribution may exist around both U sites. 
A consequence of the reduced symmetry (compared to the very high symmetry observed, for example, in \textit{fcc} UN) is that the local configuration around each U atom may not have an inversion center. 
This may be seen from Fig. 1 of Ref. \cite{Moon1968}. 
Instead of an eightfold coordination of nitrogen about each actinide ion, there are only 6 for both the C$_{3i}$ and C$_2$ sites. 
The positions of the nitrogen are also marked in Ref. \cite{Troc1975}, Fig. 7, and these figures also show the non-centrosymmetric local coordination around each U atom. 
Although such a lack of inversion center may be related to the physical reason the quadrupoles exist, we should emphasize that the symmetry elements alone determine whether this phenomenon can be observed.

\begin{table*}[ht]
\centering
\caption{Table of structure factors for the first 11 charge reflections in the cubic U$_2$N$_3$ structure; space group \# 206. The cubic lattice parameter used is $a$ = 10.69 \AA{}. The quantities SF U$_1$ and U$_2$ are the trigonometric structure factors for the uranium atoms for each reflection. No nitrogen atoms are included (there are 48 in the unit cell), and no scattering factor for the U atoms. The reflections in the first column with an “x” indicate that the reflection is forbidden $-$ group B; where the group A is with all the atoms in phase, and thus a large intensity. The groups are discussed in the text. In the Observation (Obs.) column, Ma and Ch signify magnetic and charge energy profiles, see Fig. \ref{fig:en} lower panel. The final column gives the origin of the ATS structure factor, see text. Reflections indices here follow the convention for cubic systems, $h$ \textgreater  $k$ \textgreater  $l$, whereas figures correspond to the observed reflections with the specular direction defined as \hkl[001].}
\label{table:sf}
\begin{ruledtabular}
\begin{tabular}{cccccccccc}
\multicolumn{1}{c}{\hkl(hkl)} & \multicolumn{1}{c}{$\left|Q\right|$ (\AA{}$^{-1}$)} & \multicolumn{1}{c}{SF U$_1$} & \multicolumn{1}{c}{SF U$_2$} & \multicolumn{1}{c}{Total SF} & \multicolumn{1}{c}{Group} & \multicolumn{1}{c}{Observation} & \multicolumn{2}{c}{ATS} \\ \hline
\hkl(000)                    & 0                                                                                                & 8                            & 24                           & 32                           & A                         &                                 &            &            \\
\hkl(110)x         & 0.831                                                                                            & 0.00                         & 0.00                         & 0.00                         & B                         &                                 & U$_1$      & U$_2$      \\
\hkl(200)                     & 1.176                                                                                            & -8.00                        & 7.75                         & -0.25                        & C                         & Ma                              &            & U$_2$      \\
\hkl(211)                     & 1.440                                                                                            & 0.00                         & 1.99                         & 1.99                         & D                         & Ch                              & U$_1$      & U$_2$      \\
\hkl(220)                     & 1.662                                                                                            & 8.00                         & -8.00                        & 0.00                         & E                         & Ma                              &            & U$_2$      \\
\hkl(310)x         & 1.859                                                                                            & 0.00                         & 0.00                         & 0.00                         & B                         & Ma                              & U$_1$      & U$_2$      \\
\hkl(222)                     & 2.036                                                                                            & -8.00                        & -23.25                        & -31.25                       & A                         & Ch                              &            &            \\
\hkl(321)                     & 2.199                                                                                            & 0.00                         & -1.99                        & -1.99                        & D                         &                                 & U$_1$      & U$_2$      \\
\hkl(400)                     & 2.351                                                                                            & 8.00                         & 23.01                        & 31.01                        & A                         & Ch                              &            &            \\
\hkl(330)x         & 2.494                                                                                            & 0.00                         & 0.00                         & 0.00                         & B                         &                                 & U$_1$      & U$_2$      \\
\hkl(411)                     & 2.494                                                                                            & 0.00                         & 3.85                         & 3.85                         & D                         & Ch                              & U$_1$      & U$_2$      \\
\hkl(420)                     & 2.494                                                                                            & -8.00                        & 6.76                         & -1.24                        & C                         &                                 &            & U$_2$     
\end{tabular}
\end{ruledtabular}
\end{table*}

The full scattering factor may be written
\begin{equation}
			f = f_0 + f' + i f''
\end{equation}
where the first two terms are real, and the last term is imaginary. 
The last two terms are zero unless the x-ray wavelength is near an absorption edge. 
Normally, if the environment of the atom is highly symmetric, the energy dependence of the charge scattering will resemble the well-known dispersive shape, as shown in Fig. \ref{fig:en}(a). 
However, if the spatial dependence of the electronic distribution has an \textit{aspherical} contribution from the quadrupoles, then, with the energy close to an absorption edge this aspherical distribution will couple to $f''$. 
Of course, the spherically symmetric part will always be present, so the much smaller asymmetric part cannot be observed unless the spherically symmetric part cancels. 
U$_2$N$_3$ gives a good illustration of this, as shown in the lower part of Fig. \ref{fig:en}.

Figure \ref{fig:en}(c) shows the energy dependence of various charge reflections allowed in the \textit{bcc} structure of U$_2$N$_3$. 
With reference to the groups listed in Table \ref{table:sf}, we see that groups B, C, and E sense the imaginary term $f''$, whereas group D has the real part, $f_0 + f'$, energy dependence. 
Group A reflections have all contributions in phase and are not sensitive to the aspherical distribution. 
Group B are forbidden reflections, group C would be zero if $x = 0$ for the U$_2$ atom, and are thus weak, and group E have the spherical contribution from the two U sites cancelling, and are thus also weak. 
The scattering from the nitrogen atoms, present in all reflections, except group B, is weak compared with any scattering from uranium, so allows the ATS contribution to be also observed in groups C and E. 

Previously, ATS scattering has mainly been observed at transition metal $K$ edges \cite{Collins2001,Kokubun2012}. 
However, disentangling the physics from such $K$-edge measurements is difficult, as the $K$ edge corresponds to 1$s$ to n$p$ dipole transitions, where n is the first partially filled $p$ shell, and higher-order transitions, (for example, the quadrupole transition is 1$s$ to n$d$) can contribute. 
In our case, we know that the transitions are dipole in nature and that the aspherical part of the electron density involves the 5$f$ electrons, because we observe the effect at the U $M_4$ edge. 
To our knowledge no such comparable observation involving 5$f$ electrons has been reported previously. 
To verify that this scattering is truly ATS we have performed an azimuthal scan (not shown) on the \hkl(002) reflection, and we have also shown that the ATS scattering of all reflections is independent of temperature. 
The magnetically-driven ordering of quadrupoles, such as is found in UO$_2$ and NpO$_2$ \cite{Santini2009}, would have a different azimuthal and energy dependence, and are dependent on temperature, with no signal for T \textgreater T$_N$.

Table \ref{table:sf} (final column) shows that there are contributions of the ATS from all reflections from the U$_2$ atom, but groups C and E have \textit{no} contribution from the U$_1$ atom. 
Since we see ATS scattering from group C and E reflections, the aspherical distribution must be present around the U$_2$ site. 
Although we cannot exclude its presence around U$_1$, the fact that the \hkl(022) (group E) and \hkl(013) (group B) contributions, in Fig. \ref{fig:en} (c), are of the \textit{same} magnitude, suggests that any U$_1$ aspherical contribution is very small, as the \hkl(013) has contributions from both U$_1$ and U$_2$, whereas the \hkl(022) has contributions from only U$_2$. 
These two reflections have a Lorentz factor ($1 / sin 2\theta$) of $\sim$ 1.2, and are close in $\left|Q\right|$. 
The Lorentz factor for the \hkl(002) reflection is 1.71 so it should be stronger than reflections at higher Q. 
The overall Q-dependence for intensities from this dipole transition is still subject of discussion \cite{Hill1996,Watson1998}.

\section{Conclusions}

\subsection{UN}

Early work on UN assumed that the electronic configuration was 5$f^2$ and that the 5$f$ electrons were localized. 
Using well-known crystal-field theory, many of the experimental results could be explained on this basis. 
However, the first neutron inelastic scattering experiments on UN in 1974 failed to find any distinct crystal-field levels \cite{Wedgwood1974}, and no evidence has been found for such levels in more recent experiments \cite{Aczel2012a,Lin2014}, so that this theory does not seem immediately relevant. 
We assume that the crystal-field levels are heavily damped by the interactions between the 5$f$ and conduction-electron states. 
The advent of band theory changed the perspective on even some of the earlier experimental results, in that the orbital moment could be incorporated into such theories \cite{Brooks1983}. 
Two important experiments, both on single crystals, added further weight to the idea that the 5$f$ electrons are itinerant in UN, firstly, the neutron inelastic scattering \cite{Holden1984}, and, secondly, the measurement of angular resolved photoemission spectroscopy \cite{Fujimori2012,Fujimori2016}. 

Although, some properties of UN might still be described with localized 5$f$ states, giving rise to the idea of duality in UN \cite{Troc2016}, the weight of evidence points to the best approach being one with band (itinerant) 5$f$ electrons. 
The theoretical calculations mentioned earlier \cite{Yin2011,Szpunar2014} both use such assumptions, with the 5$f$ states numbering between two and three 5$f$ electrons. 
On the other hand, these calculations do not reproduce the correct (as measured by experiment) AF magnetic moment in UN, and no effort that we are aware of has been made to calculate the ground-state antiferromagnetic state and associated moment - this being an excellent test as to whether the assumed electronic structure is a true representation of the material. 
When such calculations are made, we can see which of the 1\textbf{k} or 3\textbf{k} magnetic configurations is the more stable state. 
More complicated is to replicate the \textit{expansion} of the unit cell below T$_N$, and the effects of the magnetic fields on the system \cite{Troc2016,Shrestha2017}, which depend crucially on the AF ground state. 
These large effects with applied magnetic field \cite{Gorbunov2019} reflect a strong coupling between spin, electronic, and lattice degrees of freedom.

\subsection{U$_2$N$_3$}

Although this material is closely linked to UN, very few investigations of its electronic structure have been reported.
The crystal structure and its synthesis are well recorded. 
We have found that our sample orders at 73.5(2) K, with a lattice parameter of 10.80 \AA{}. 
The magnetic wave-vector is \textbf{q} = \hkl<001>, which is the same as that found for Yb$_2$O$_3$ \cite{Moon1968}. 
The magnetic configuration may well be non-collinear, but we cannot determine this from our measurements. 

In terms of valencies on the individual atomic sites, this is a question that might be answered if we knew whether both sites ordered magnetically. 
We cannot answer that conclusively with the results of the x-ray experiments reported here. 
From the U 4$f_{7/2}$ spectra reported for U$_2$N$_3$ in \cite{LawrenceBright2018} there is a shift in the weight of the spectra towards a higher oxidation state than reported for UN. 
If we assume that the majority 24 U$_2$ sites are 3+ (i.e. approximately the same as in UN), then charge compensation (taking N$^{3-}$) suggests the 8 U$_1$ sites might well be of a higher oxidation state, perhaps even U$^{6+}$. 
Since U$^{6+}$ is soluble in water and highly reactive, this would explain why the U$_2$N$_3$ is more reactive (in H$_2$O$_2$) than either UO$_2$ or UN, as reported in Ref. \cite{LawrenceBright2019}. 
U$^{6+}$ would have no associated 5$f$ electrons, so would \textit{not} order magnetically, nor should there be any aspherical resonant scattering from U$_1$ as suggested by the observations in Fig.\ref{fig:en} (c). 
This would be consistent with our observation that the U$_1$ atom probably has zero (or at least small) aspherical contribution to the ATS scattering.

Of course, such counting of charges is certainly too simple an approach in a material that is certainly a semimetal, and we welcome some theoretical interest given that U$_2$N$_3$ is always found in conjunction with (and especially at the surface of) UN.

Similarly, a theoretical investigation should be able to throw light on the ATS scattering reported for U$_2$N$_3$ shown in Fig. \ref{fig:en} and believed to be associated only with the U$_2$ atom. 
For example, it seems probable that this is related to the hybridization between the U 5$f$ states and the N 2$p$ states and directly represents 5$f$ covalency. 
Such effects might be a widespread property of actinide compounds, but for its observation requires the special conditions afforded by forbidden reflections in the bixbyite structure \cite{Kokubun2012}, which exists for U$_2$N$_3$, but not for UN.

\section*{Acknowledgements}

We acknowledge Diamond Light Source for time on Beamline I16 under Proposal 20776 and funding from EPSRC grant 1652612.

\bibliography{rsprefs}

\begin{thebibliography}{51}
\expandafter\ifx\csname natexlab\endcsname\relax\def\natexlab#1{#1}\fi
\expandafter\ifx\csname bibnamefont\endcsname\relax
  \def\bibnamefont#1{#1}\fi
\expandafter\ifx\csname bibfnamefont\endcsname\relax
  \def\bibfnamefont#1{#1}\fi
\expandafter\ifx\csname citenamefont\endcsname\relax
  \def\citenamefont#1{#1}\fi
\expandafter\ifx\csname url\endcsname\relax
  \def\url#1{\texttt{#1}}\fi
\expandafter\ifx\csname urlprefix\endcsname\relax\def\urlprefix{URL }\fi
\providecommand{\bibinfo}[2]{#2}
\providecommand{\eprint}[2][]{\url{#2}}

\bibitem[{\citenamefont{Kurosaki et~al.}(2000)\citenamefont{Kurosaki, Yano,
  Yamada, Uno, and Yamanaka}}]{Kurosaki2000}
\bibinfo{author}{\bibfnamefont{K.}~\bibnamefont{Kurosaki}},
  \bibinfo{author}{\bibfnamefont{K.}~\bibnamefont{Yano}},
  \bibinfo{author}{\bibfnamefont{K.}~\bibnamefont{Yamada}},
  \bibinfo{author}{\bibfnamefont{M.}~\bibnamefont{Uno}}, \bibnamefont{and}
  \bibinfo{author}{\bibfnamefont{S.}~\bibnamefont{Yamanaka}},
  \bibinfo{journal}{Journal of Alloys and Compounds}
  \textbf{\bibinfo{volume}{311}}, \bibinfo{pages}{305} (\bibinfo{year}{2000}),
  ISSN \bibinfo{issn}{09258388}.

\bibitem[{\citenamefont{Ronchi et~al.}(2004)\citenamefont{Ronchi, Sheindlin,
  Staicu, and Kinoshita}}]{Ronchi2004}
\bibinfo{author}{\bibfnamefont{C.}~\bibnamefont{Ronchi}},
  \bibinfo{author}{\bibfnamefont{M.}~\bibnamefont{Sheindlin}},
  \bibinfo{author}{\bibfnamefont{D.}~\bibnamefont{Staicu}}, \bibnamefont{and}
  \bibinfo{author}{\bibfnamefont{M.}~\bibnamefont{Kinoshita}},
  \bibinfo{journal}{Journal of Nuclear Materials}
  \textbf{\bibinfo{volume}{327}}, \bibinfo{pages}{58} (\bibinfo{year}{2004}),
  ISSN \bibinfo{issn}{00223115}.

\bibitem[{\citenamefont{Yin et~al.}(2011)\citenamefont{Yin, Kutepov, Haule,
  Kotliar, Savrasov, and Pickett}}]{Yin2011}
\bibinfo{author}{\bibfnamefont{Q.}~\bibnamefont{Yin}},
  \bibinfo{author}{\bibfnamefont{A.}~\bibnamefont{Kutepov}},
  \bibinfo{author}{\bibfnamefont{K.}~\bibnamefont{Haule}},
  \bibinfo{author}{\bibfnamefont{G.}~\bibnamefont{Kotliar}},
  \bibinfo{author}{\bibfnamefont{S.~Y.} \bibnamefont{Savrasov}},
  \bibnamefont{and} \bibinfo{author}{\bibfnamefont{W.~E.}
  \bibnamefont{Pickett}}, \bibinfo{journal}{Physical Review B - Condensed
  Matter and Materials Physics} \textbf{\bibinfo{volume}{84}},
  \bibinfo{pages}{1} (\bibinfo{year}{2011}), ISSN \bibinfo{issn}{10980121},
  \eprint{1012.2412}.

\bibitem[{\citenamefont{Szpunar and Szpunar}(2014)}]{Szpunar2014}
\bibinfo{author}{\bibfnamefont{B.}~\bibnamefont{Szpunar}} \bibnamefont{and}
  \bibinfo{author}{\bibfnamefont{J.~A.} \bibnamefont{Szpunar}},
  \bibinfo{journal}{International Journal of Nuclear Energy}
  \textbf{\bibinfo{volume}{2014}} (\bibinfo{year}{2014}), ISSN
  \bibinfo{issn}{2356-7066}.

\bibitem[{\citenamefont{{Lawrence Bright} et~al.}(2018)\citenamefont{{Lawrence
  Bright}, Rennie, Cattelan, Fox, Goddard, and Springell}}]{LawrenceBright2018}
\bibinfo{author}{\bibfnamefont{E.}~\bibnamefont{{Lawrence Bright}}},
  \bibinfo{author}{\bibfnamefont{S.}~\bibnamefont{Rennie}},
  \bibinfo{author}{\bibfnamefont{M.}~\bibnamefont{Cattelan}},
  \bibinfo{author}{\bibfnamefont{N.~A.} \bibnamefont{Fox}},
  \bibinfo{author}{\bibfnamefont{D.~T.} \bibnamefont{Goddard}},
  \bibnamefont{and}
  \bibinfo{author}{\bibfnamefont{R.}~\bibnamefont{Springell}},
  \bibinfo{journal}{Thin Solid Films} \textbf{\bibinfo{volume}{661}},
  \bibinfo{pages}{71} (\bibinfo{year}{2018}), ISSN \bibinfo{issn}{00406090},
  \urlprefix\url{https://doi.org/10.1016/j.tsf.2018.07.018}.

\bibitem[{\citenamefont{{Lawrence Bright} et~al.}(2019)\citenamefont{{Lawrence
  Bright}, Rennie, Siberry, Samani, Clarke, Goddard, and
  Springell}}]{LawrenceBright2019}
\bibinfo{author}{\bibfnamefont{E.}~\bibnamefont{{Lawrence Bright}}},
  \bibinfo{author}{\bibfnamefont{S.}~\bibnamefont{Rennie}},
  \bibinfo{author}{\bibfnamefont{A.}~\bibnamefont{Siberry}},
  \bibinfo{author}{\bibfnamefont{K.}~\bibnamefont{Samani}},
  \bibinfo{author}{\bibfnamefont{K.}~\bibnamefont{Clarke}},
  \bibinfo{author}{\bibfnamefont{D.~T.} \bibnamefont{Goddard}},
  \bibnamefont{and}
  \bibinfo{author}{\bibfnamefont{R.}~\bibnamefont{Springell}},
  \bibinfo{journal}{Journal of Nuclear Materials}
  \textbf{\bibinfo{volume}{518}}, \bibinfo{pages}{202} (\bibinfo{year}{2019}),
  ISSN \bibinfo{issn}{00223115},
  \urlprefix\url{https://doi.org/10.1016/j.jnucmat.2019.03.006}.

\bibitem[{\citenamefont{Curry}(1965)}]{Curry1965}
\bibinfo{author}{\bibfnamefont{N.~A.} \bibnamefont{Curry}},
  \bibinfo{journal}{Proceeding of the Physical Society}
  \textbf{\bibinfo{volume}{86}}, \bibinfo{pages}{1193} (\bibinfo{year}{1965}),
  ISSN \bibinfo{issn}{03701328}.

\bibitem[{\citenamefont{Tro{\'{c}} et~al.}(2016)\citenamefont{Tro{\'{c}},
  Samsel-Czeka{\l}a, Pikul, Andreev, Gorbunov, Skourski, and
  Sznajd}}]{Troc2016}
\bibinfo{author}{\bibfnamefont{R.}~\bibnamefont{Tro{\'{c}}}},
  \bibinfo{author}{\bibfnamefont{M.}~\bibnamefont{Samsel-Czeka{\l}a}},
  \bibinfo{author}{\bibfnamefont{A.}~\bibnamefont{Pikul}},
  \bibinfo{author}{\bibfnamefont{A.~V.} \bibnamefont{Andreev}},
  \bibinfo{author}{\bibfnamefont{D.~I.} \bibnamefont{Gorbunov}},
  \bibinfo{author}{\bibfnamefont{Y.}~\bibnamefont{Skourski}}, \bibnamefont{and}
  \bibinfo{author}{\bibfnamefont{J.}~\bibnamefont{Sznajd}},
  \bibinfo{journal}{Physical Review B} \textbf{\bibinfo{volume}{94}},
  \bibinfo{pages}{224415} (\bibinfo{year}{2016}), ISSN
  \bibinfo{issn}{24699969}.

\bibitem[{\citenamefont{Rundle et~al.}(1948)\citenamefont{Rundle, Baenziger,
  Wilson, and McDonald}}]{Rundle1948}
\bibinfo{author}{\bibfnamefont{R.~E.} \bibnamefont{Rundle}},
  \bibinfo{author}{\bibfnamefont{N.~C.} \bibnamefont{Baenziger}},
  \bibinfo{author}{\bibfnamefont{A.~S.} \bibnamefont{Wilson}},
  \bibnamefont{and} \bibinfo{author}{\bibfnamefont{R.~A.}
  \bibnamefont{McDonald}}, \bibinfo{journal}{J. Am. Chem. Soc.}
  \textbf{\bibinfo{volume}{70}}, \bibinfo{pages}{99} (\bibinfo{year}{1948}).

\bibitem[{\citenamefont{Troc}(1975)}]{Troc1975}
\bibinfo{author}{\bibfnamefont{R.}~\bibnamefont{Troc}},
  \bibinfo{journal}{Journal of Solid State Chemistry}
  \textbf{\bibinfo{volume}{13}}, \bibinfo{pages}{14} (\bibinfo{year}{1975}),
  ISSN \bibinfo{issn}{00224596}.

\bibitem[{\citenamefont{Black et~al.}(2001)\citenamefont{Black, Miserque,
  Gouder, Havela, Rebizant, and Wastin}}]{Black2001}
\bibinfo{author}{\bibfnamefont{L.}~\bibnamefont{Black}},
  \bibinfo{author}{\bibfnamefont{F.}~\bibnamefont{Miserque}},
  \bibinfo{author}{\bibfnamefont{T.}~\bibnamefont{Gouder}},
  \bibinfo{author}{\bibfnamefont{L.}~\bibnamefont{Havela}},
  \bibinfo{author}{\bibfnamefont{J.}~\bibnamefont{Rebizant}}, \bibnamefont{and}
  \bibinfo{author}{\bibfnamefont{F.}~\bibnamefont{Wastin}},
  \bibinfo{journal}{Journal of Alloys and Compounds}
  \textbf{\bibinfo{volume}{315}}, \bibinfo{pages}{36} (\bibinfo{year}{2001}),
  ISSN \bibinfo{issn}{09258388}.

\bibitem[{\citenamefont{Rafaja et~al.}(2005)\citenamefont{Rafaja, Havela,
  Ku{\v{z}}el, Wastin, Colineau, and Gouder}}]{Rafaja2005}
\bibinfo{author}{\bibfnamefont{D.}~\bibnamefont{Rafaja}},
  \bibinfo{author}{\bibfnamefont{L.}~\bibnamefont{Havela}},
  \bibinfo{author}{\bibfnamefont{R.}~\bibnamefont{Ku{\v{z}}el}},
  \bibinfo{author}{\bibfnamefont{F.}~\bibnamefont{Wastin}},
  \bibinfo{author}{\bibfnamefont{E.}~\bibnamefont{Colineau}}, \bibnamefont{and}
  \bibinfo{author}{\bibfnamefont{T.}~\bibnamefont{Gouder}},
  \bibinfo{journal}{Journal of Alloys and Compounds}
  \textbf{\bibinfo{volume}{386}}, \bibinfo{pages}{87} (\bibinfo{year}{2005}),
  ISSN \bibinfo{issn}{09258388}.

\bibitem[{\citenamefont{Zhang et~al.}(2010)\citenamefont{Zhang, Meng, Xu, and
  Zhang}}]{Zhang2010}
\bibinfo{author}{\bibfnamefont{Y.}~\bibnamefont{Zhang}},
  \bibinfo{author}{\bibfnamefont{D.}~\bibnamefont{Meng}},
  \bibinfo{author}{\bibfnamefont{Q.}~\bibnamefont{Xu}}, \bibnamefont{and}
  \bibinfo{author}{\bibfnamefont{Y.}~\bibnamefont{Zhang}},
  \bibinfo{journal}{Journal of Nuclear Materials}
  \textbf{\bibinfo{volume}{397}}, \bibinfo{pages}{31} (\bibinfo{year}{2010}),
  ISSN \bibinfo{issn}{00223115},
  \urlprefix\url{http://dx.doi.org/10.1016/j.jnucmat.2009.12.002}.

\bibitem[{\citenamefont{Long et~al.}(2016)\citenamefont{Long, Luo, Lu, Hu, Liu,
  and Lai}}]{Long2016}
\bibinfo{author}{\bibfnamefont{Z.}~\bibnamefont{Long}},
  \bibinfo{author}{\bibfnamefont{L.}~\bibnamefont{Luo}},
  \bibinfo{author}{\bibfnamefont{Y.}~\bibnamefont{Lu}},
  \bibinfo{author}{\bibfnamefont{Y.}~\bibnamefont{Hu}},
  \bibinfo{author}{\bibfnamefont{K.}~\bibnamefont{Liu}}, \bibnamefont{and}
  \bibinfo{author}{\bibfnamefont{X.}~\bibnamefont{Lai}},
  \bibinfo{journal}{Journal of Alloys and Compounds}
  \textbf{\bibinfo{volume}{664}}, \bibinfo{pages}{745} (\bibinfo{year}{2016}),
  ISSN \bibinfo{issn}{09258388},
  \urlprefix\url{http://dx.doi.org/10.1016/j.jallcom.2016.01.013}.

\bibitem[{\citenamefont{Wang et~al.}(2016)\citenamefont{Wang, Long, Bin, Yang,
  Pan, Li, Luo, Hu, and Liu}}]{Wang2016}
\bibinfo{author}{\bibfnamefont{X.}~\bibnamefont{Wang}},
  \bibinfo{author}{\bibfnamefont{Z.}~\bibnamefont{Long}},
  \bibinfo{author}{\bibfnamefont{R.}~\bibnamefont{Bin}},
  \bibinfo{author}{\bibfnamefont{R.}~\bibnamefont{Yang}},
  \bibinfo{author}{\bibfnamefont{Q.}~\bibnamefont{Pan}},
  \bibinfo{author}{\bibfnamefont{F.}~\bibnamefont{Li}},
  \bibinfo{author}{\bibfnamefont{L.}~\bibnamefont{Luo}},
  \bibinfo{author}{\bibfnamefont{Y.}~\bibnamefont{Hu}}, \bibnamefont{and}
  \bibinfo{author}{\bibfnamefont{K.}~\bibnamefont{Liu}},
  \bibinfo{journal}{Inorganic Chemistry} \textbf{\bibinfo{volume}{55}},
  \bibinfo{pages}{10835} (\bibinfo{year}{2016}).

\bibitem[{\citenamefont{Lu et~al.}(2016)\citenamefont{Lu, Li, Hu, Xiao, Bai,
  Zhang, Luo, Liu, and Liu}}]{Lu2016}
\bibinfo{author}{\bibfnamefont{L.}~\bibnamefont{Lu}},
  \bibinfo{author}{\bibfnamefont{F.}~\bibnamefont{Li}},
  \bibinfo{author}{\bibfnamefont{Y.}~\bibnamefont{Hu}},
  \bibinfo{author}{\bibfnamefont{H.}~\bibnamefont{Xiao}},
  \bibinfo{author}{\bibfnamefont{B.}~\bibnamefont{Bai}},
  \bibinfo{author}{\bibfnamefont{Y.}~\bibnamefont{Zhang}},
  \bibinfo{author}{\bibfnamefont{L.}~\bibnamefont{Luo}},
  \bibinfo{author}{\bibfnamefont{J.}~\bibnamefont{Liu}}, \bibnamefont{and}
  \bibinfo{author}{\bibfnamefont{K.}~\bibnamefont{Liu}},
  \bibinfo{journal}{Journal of Nuclear Materials}
  \textbf{\bibinfo{volume}{480}}, \bibinfo{pages}{189} (\bibinfo{year}{2016}),
  ISSN \bibinfo{issn}{00223115}.

\bibitem[{\citenamefont{Collins et~al.}(2010)\citenamefont{Collins, Bombardi,
  Marshall, Williams, Barlow, Day, Pearson, Woolliscroft, Walton, Beutier
  et~al.}}]{Collins2010}
\bibinfo{author}{\bibfnamefont{S.~P.} \bibnamefont{Collins}},
  \bibinfo{author}{\bibfnamefont{A.}~\bibnamefont{Bombardi}},
  \bibinfo{author}{\bibfnamefont{A.~R.} \bibnamefont{Marshall}},
  \bibinfo{author}{\bibfnamefont{J.~H.} \bibnamefont{Williams}},
  \bibinfo{author}{\bibfnamefont{G.}~\bibnamefont{Barlow}},
  \bibinfo{author}{\bibfnamefont{A.~G.} \bibnamefont{Day}},
  \bibinfo{author}{\bibfnamefont{M.~R.} \bibnamefont{Pearson}},
  \bibinfo{author}{\bibfnamefont{R.~J.} \bibnamefont{Woolliscroft}},
  \bibinfo{author}{\bibfnamefont{R.~D.} \bibnamefont{Walton}},
  \bibinfo{author}{\bibfnamefont{G.}~\bibnamefont{Beutier}},
  \bibnamefont{et~al.}, \bibinfo{journal}{AIP Conference Proceedings}
  \textbf{\bibinfo{volume}{1234}}, \bibinfo{pages}{303} (\bibinfo{year}{2010}),
  ISSN \bibinfo{issn}{0094243X}.

\bibitem[{\citenamefont{Marples}(1970)}]{Marples1970}
\bibinfo{author}{\bibfnamefont{J.~A.} \bibnamefont{Marples}},
  \bibinfo{journal}{Journal of Physics and Chemistry of Solids}
  \textbf{\bibinfo{volume}{31}}, \bibinfo{pages}{2431} (\bibinfo{year}{1970}),
  ISSN \bibinfo{issn}{00223697}.

\bibitem[{\citenamefont{Marples et~al.}(1975)\citenamefont{Marples, Sampson,
  Wedgwood, and Kuznietz}}]{Marples1975}
\bibinfo{author}{\bibfnamefont{J.~A.} \bibnamefont{Marples}},
  \bibinfo{author}{\bibfnamefont{C.~F.} \bibnamefont{Sampson}},
  \bibinfo{author}{\bibfnamefont{F.~A.} \bibnamefont{Wedgwood}},
  \bibnamefont{and} \bibinfo{author}{\bibfnamefont{M.}~\bibnamefont{Kuznietz}},
  \bibinfo{journal}{Journal of Physics C: Solid State Physics}
  \textbf{\bibinfo{volume}{8}}, \bibinfo{pages}{708} (\bibinfo{year}{1975}),
  ISSN \bibinfo{issn}{00223719}.

\bibitem[{\citenamefont{Doorn and Plessis}(1977)}]{Doorn1977}
\bibinfo{author}{\bibfnamefont{C.~F.} \bibnamefont{Doorn}} \bibnamefont{and}
  \bibinfo{author}{\bibfnamefont{P.~D.~V.} \bibnamefont{Plessis}},
  \bibinfo{journal}{Journal of Low Temperature Physics}
  \textbf{\bibinfo{volume}{28}}, \bibinfo{pages}{391} (\bibinfo{year}{1977}),
  ISSN \bibinfo{issn}{0022-2291},
  \urlprefix\url{http://www.springerlink.com/index/10.1007/BF00668226}.

\bibitem[{\citenamefont{Shrestha et~al.}(2017)\citenamefont{Shrestha, Antonio,
  Jaime, Harrison, Mast, Safarik, Durakiewicz, Griveau, and
  Gofryk}}]{Shrestha2017}
\bibinfo{author}{\bibfnamefont{K.}~\bibnamefont{Shrestha}},
  \bibinfo{author}{\bibfnamefont{D.}~\bibnamefont{Antonio}},
  \bibinfo{author}{\bibfnamefont{M.}~\bibnamefont{Jaime}},
  \bibinfo{author}{\bibfnamefont{N.}~\bibnamefont{Harrison}},
  \bibinfo{author}{\bibfnamefont{D.~S.} \bibnamefont{Mast}},
  \bibinfo{author}{\bibfnamefont{D.}~\bibnamefont{Safarik}},
  \bibinfo{author}{\bibfnamefont{T.}~\bibnamefont{Durakiewicz}},
  \bibinfo{author}{\bibfnamefont{J.}~\bibnamefont{Griveau}}, \bibnamefont{and}
  \bibinfo{author}{\bibfnamefont{K.}~\bibnamefont{Gofryk}},
  \bibinfo{journal}{Scientific Reports} \textbf{\bibinfo{volume}{7}},
  \bibinfo{pages}{6642} (\bibinfo{year}{2017}), ISSN \bibinfo{issn}{2045-2322},
  \urlprefix\url{http://dx.doi.org/10.1038/s41598-017-06154-7}.

\bibitem[{\citenamefont{Rossat-Mignod et~al.}(1980)\citenamefont{Rossat-Mignod,
  Burlet, Quezel, and Vogt}}]{Rossat-Mignod1980}
\bibinfo{author}{\bibfnamefont{J.}~\bibnamefont{Rossat-Mignod}},
  \bibinfo{author}{\bibfnamefont{P.}~\bibnamefont{Burlet}},
  \bibinfo{author}{\bibfnamefont{S.}~\bibnamefont{Quezel}}, \bibnamefont{and}
  \bibinfo{author}{\bibfnamefont{O.}~\bibnamefont{Vogt}},
  \bibinfo{journal}{Physica B+C} \textbf{\bibinfo{volume}{102}},
  \bibinfo{pages}{237} (\bibinfo{year}{1980}), ISSN \bibinfo{issn}{03784363}.

\bibitem[{\citenamefont{Knott et~al.}(1980)\citenamefont{Knott, Lander,
  Mueller, and Vogt}}]{Knott1980}
\bibinfo{author}{\bibfnamefont{H.~W.} \bibnamefont{Knott}},
  \bibinfo{author}{\bibfnamefont{G.~H.} \bibnamefont{Lander}},
  \bibinfo{author}{\bibfnamefont{M.~H.} \bibnamefont{Mueller}},
  \bibnamefont{and} \bibinfo{author}{\bibfnamefont{O.}~\bibnamefont{Vogt}},
  \bibinfo{journal}{Physical Review B} \textbf{\bibinfo{volume}{21}},
  \bibinfo{pages}{4159} (\bibinfo{year}{1980}).

\bibitem[{\citenamefont{Blackburn et~al.}(2005)\citenamefont{Blackburn,
  Caciuffo, Magnani, Santini, Brown, Enderle, and Lander}}]{Blackburn2005}
\bibinfo{author}{\bibfnamefont{E.}~\bibnamefont{Blackburn}},
  \bibinfo{author}{\bibfnamefont{R.}~\bibnamefont{Caciuffo}},
  \bibinfo{author}{\bibfnamefont{N.}~\bibnamefont{Magnani}},
  \bibinfo{author}{\bibfnamefont{P.}~\bibnamefont{Santini}},
  \bibinfo{author}{\bibfnamefont{P.~J.} \bibnamefont{Brown}},
  \bibinfo{author}{\bibfnamefont{M.}~\bibnamefont{Enderle}}, \bibnamefont{and}
  \bibinfo{author}{\bibfnamefont{G.~H.} \bibnamefont{Lander}},
  \bibinfo{journal}{Physical Review B - Condensed Matter and Materials Physics}
  \textbf{\bibinfo{volume}{72}}, \bibinfo{pages}{184411}
  (\bibinfo{year}{2005}), ISSN \bibinfo{issn}{10980121}.

\bibitem[{\citenamefont{Magnani et~al.}(2010)\citenamefont{Magnani, Caciuffo,
  Lander, Hiess, and Regnault}}]{Magnani2010}
\bibinfo{author}{\bibfnamefont{N.}~\bibnamefont{Magnani}},
  \bibinfo{author}{\bibfnamefont{R.}~\bibnamefont{Caciuffo}},
  \bibinfo{author}{\bibfnamefont{G.~H.} \bibnamefont{Lander}},
  \bibinfo{author}{\bibfnamefont{A.}~\bibnamefont{Hiess}}, \bibnamefont{and}
  \bibinfo{author}{\bibfnamefont{L.~P.} \bibnamefont{Regnault}},
  \bibinfo{journal}{Journal of Physics Condensed Matter}
  \textbf{\bibinfo{volume}{22}}, \bibinfo{pages}{116002}
  (\bibinfo{year}{2010}).

\bibitem[{\citenamefont{Hill and Mcmorrow}(1996)}]{Hill1996}
\bibinfo{author}{\bibfnamefont{J.~P.} \bibnamefont{Hill}} \bibnamefont{and}
  \bibinfo{author}{\bibfnamefont{D.~F.} \bibnamefont{Mcmorrow}},
  \bibinfo{journal}{Acta Crystallographica Section A: Foundations of
  Crystallography} \textbf{\bibinfo{volume}{52}}, \bibinfo{pages}{236}
  (\bibinfo{year}{1996}), ISSN \bibinfo{issn}{01087673}.

\bibitem[{\citenamefont{Cross et~al.}(1998)\citenamefont{Cross, Newville, Rehr,
  Sorensen, Bouldin, Watson, Gouder, Lander, and Bell}}]{Cross1998}
\bibinfo{author}{\bibfnamefont{J.~O.} \bibnamefont{Cross}},
  \bibinfo{author}{\bibfnamefont{M.}~\bibnamefont{Newville}},
  \bibinfo{author}{\bibfnamefont{J.~J.} \bibnamefont{Rehr}},
  \bibinfo{author}{\bibfnamefont{L.~B.} \bibnamefont{Sorensen}},
  \bibinfo{author}{\bibfnamefont{C.~E.} \bibnamefont{Bouldin}},
  \bibinfo{author}{\bibfnamefont{G.}~\bibnamefont{Watson}},
  \bibinfo{author}{\bibfnamefont{T.}~\bibnamefont{Gouder}},
  \bibinfo{author}{\bibfnamefont{G.~H.} \bibnamefont{Lander}},
  \bibnamefont{and} \bibinfo{author}{\bibfnamefont{M.~I.} \bibnamefont{Bell}},
  \bibinfo{journal}{Physical Review B - Condensed Matter and Materials Physics}
  \textbf{\bibinfo{volume}{58}}, \bibinfo{pages}{11215} (\bibinfo{year}{1998}),
  ISSN \bibinfo{issn}{1550235X}.

\bibitem[{\citenamefont{Watson et~al.}(1998)\citenamefont{Watson, Nuttall,
  Forgan, and Perry}}]{Watson1998}
\bibinfo{author}{\bibfnamefont{D.}~\bibnamefont{Watson}},
  \bibinfo{author}{\bibfnamefont{W.}~\bibnamefont{Nuttall}},
  \bibinfo{author}{\bibfnamefont{E.}~\bibnamefont{Forgan}}, \bibnamefont{and}
  \bibinfo{author}{\bibfnamefont{S.}~\bibnamefont{Perry}},
  \bibinfo{journal}{Physical Review B - Condensed Matter and Materials Physics}
  \textbf{\bibinfo{volume}{57}}, \bibinfo{pages}{R8095} (\bibinfo{year}{1998}),
  ISSN \bibinfo{issn}{1550235X}.

\bibitem[{\citenamefont{Holden et~al.}(1982)\citenamefont{Holden, Buyers,
  Svensson, and Lander}}]{Holden1982}
\bibinfo{author}{\bibfnamefont{T.~M.} \bibnamefont{Holden}},
  \bibinfo{author}{\bibfnamefont{J.~L.} \bibnamefont{Buyers}},
  \bibinfo{author}{\bibfnamefont{E.~C.} \bibnamefont{Svensson}},
  \bibnamefont{and} \bibinfo{author}{\bibfnamefont{G.~H.}
  \bibnamefont{Lander}}, \bibinfo{journal}{Physical Review B}
  \textbf{\bibinfo{volume}{26}} (\bibinfo{year}{1982}).

\bibitem[{\citenamefont{Lander et~al.}(1978)\citenamefont{Lander, Sinha,
  Sparlin, and Vogt}}]{Lander1978}
\bibinfo{author}{\bibfnamefont{G.~H.} \bibnamefont{Lander}},
  \bibinfo{author}{\bibfnamefont{S.~K.} \bibnamefont{Sinha}},
  \bibinfo{author}{\bibfnamefont{D.~M.} \bibnamefont{Sparlin}},
  \bibnamefont{and} \bibinfo{author}{\bibfnamefont{O.}~\bibnamefont{Vogt}},
  \bibinfo{journal}{Physical Review Letters} \textbf{\bibinfo{volume}{40}},
  \bibinfo{pages}{523} (\bibinfo{year}{1978}).

\bibitem[{\citenamefont{Bao et~al.}(2013)\citenamefont{Bao, Springell, Walker,
  Leiste, Kuebel, Prang, Nisbet, Langridge, Ward, Gouder et~al.}}]{Bao2013}
\bibinfo{author}{\bibfnamefont{Z.}~\bibnamefont{Bao}},
  \bibinfo{author}{\bibfnamefont{R.}~\bibnamefont{Springell}},
  \bibinfo{author}{\bibfnamefont{H.~C.} \bibnamefont{Walker}},
  \bibinfo{author}{\bibfnamefont{H.}~\bibnamefont{Leiste}},
  \bibinfo{author}{\bibfnamefont{K.}~\bibnamefont{Kuebel}},
  \bibinfo{author}{\bibfnamefont{R.}~\bibnamefont{Prang}},
  \bibinfo{author}{\bibfnamefont{G.}~\bibnamefont{Nisbet}},
  \bibinfo{author}{\bibfnamefont{S.}~\bibnamefont{Langridge}},
  \bibinfo{author}{\bibfnamefont{R.~C.~C.} \bibnamefont{Ward}},
  \bibinfo{author}{\bibfnamefont{T.}~\bibnamefont{Gouder}},
  \bibnamefont{et~al.}, \bibinfo{journal}{Physical Review B}
  \textbf{\bibinfo{volume}{88}}, \bibinfo{pages}{134426}
  (\bibinfo{year}{2013}), ISSN \bibinfo{issn}{1098-0121},
  \urlprefix\url{https://link.aps.org/doi/10.1103/PhysRevB.88.134426}.

\bibitem[{\citenamefont{James}(1965)}]{James1965}
\bibinfo{author}{\bibfnamefont{R.~W.} \bibnamefont{James}},
  \emph{\bibinfo{title}{{The Optical Principle of the Diffraction of X-rays}}}
  (\bibinfo{publisher}{Cornell Univ. Press.}, \bibinfo{address}{Ithaca, NY,
  USA}, \bibinfo{year}{1965}).

\bibitem[{\citenamefont{Greenberg}(1989)}]{Greenberg1989}
\bibinfo{author}{\bibfnamefont{B.}~\bibnamefont{Greenberg}},
  \bibinfo{journal}{Acta Crystallographica Section A}
  \textbf{\bibinfo{volume}{45}}, \bibinfo{pages}{238} (\bibinfo{year}{1989}),
  ISSN \bibinfo{issn}{16005724}.

\bibitem[{\citenamefont{Bernhoeft et~al.}(2004)\citenamefont{Bernhoeft, Lander,
  Longfield, Langridge, Mannix, Brown, Nuttall, Hiess, Vettier, and
  Lejay}}]{Bernhoeft2004}
\bibinfo{author}{\bibfnamefont{N.}~\bibnamefont{Bernhoeft}},
  \bibinfo{author}{\bibfnamefont{G.~H.} \bibnamefont{Lander}},
  \bibinfo{author}{\bibfnamefont{M.~J.} \bibnamefont{Longfield}},
  \bibinfo{author}{\bibfnamefont{S.}~\bibnamefont{Langridge}},
  \bibinfo{author}{\bibfnamefont{D.}~\bibnamefont{Mannix}},
  \bibinfo{author}{\bibfnamefont{S.~D.} \bibnamefont{Brown}},
  \bibinfo{author}{\bibfnamefont{W.~J.} \bibnamefont{Nuttall}},
  \bibinfo{author}{\bibfnamefont{A.}~\bibnamefont{Hiess}},
  \bibinfo{author}{\bibfnamefont{C.}~\bibnamefont{Vettier}}, \bibnamefont{and}
  \bibinfo{author}{\bibfnamefont{P.}~\bibnamefont{Lejay}},
  \bibinfo{journal}{Journal of Physics Condensed Matter}
  \textbf{\bibinfo{volume}{16}}, \bibinfo{pages}{3869} (\bibinfo{year}{2004}),
  ISSN \bibinfo{issn}{09538984}.

\bibitem[{\citenamefont{Proke{\v{s}} et~al.}(2009)\citenamefont{Proke{\v{s}},
  Lander, and Bernhoeft}}]{Prokes2009}
\bibinfo{author}{\bibfnamefont{K.}~\bibnamefont{Proke{\v{s}}}},
  \bibinfo{author}{\bibfnamefont{G.~H.} \bibnamefont{Lander}},
  \bibnamefont{and}
  \bibinfo{author}{\bibfnamefont{N.}~\bibnamefont{Bernhoeft}},
  \bibinfo{journal}{Journal of Physics Condensed Matter}
  \textbf{\bibinfo{volume}{21}}, \bibinfo{pages}{285402}
  (\bibinfo{year}{2009}).

\bibitem[{\citenamefont{Cockayne et~al.}(2013)\citenamefont{Cockayne, Levin,
  Wu, and Llobet}}]{Cockayne2013}
\bibinfo{author}{\bibfnamefont{E.}~\bibnamefont{Cockayne}},
  \bibinfo{author}{\bibfnamefont{I.}~\bibnamefont{Levin}},
  \bibinfo{author}{\bibfnamefont{H.}~\bibnamefont{Wu}}, \bibnamefont{and}
  \bibinfo{author}{\bibfnamefont{A.}~\bibnamefont{Llobet}},
  \bibinfo{journal}{Physical Review B - Condensed Matter and Materials Physics}
  \textbf{\bibinfo{volume}{87}}, \bibinfo{pages}{184413}
  (\bibinfo{year}{2013}), ISSN \bibinfo{issn}{10980121}.

\bibitem[{\citenamefont{Moon et~al.}(1968)\citenamefont{Moon, Koehler, Child,
  and Raubenheimer}}]{Moon1968}
\bibinfo{author}{\bibfnamefont{R.~M.} \bibnamefont{Moon}},
  \bibinfo{author}{\bibfnamefont{W.~C.} \bibnamefont{Koehler}},
  \bibinfo{author}{\bibfnamefont{H.~R.} \bibnamefont{Child}}, \bibnamefont{and}
  \bibinfo{author}{\bibfnamefont{L.~J.} \bibnamefont{Raubenheimer}},
  \bibinfo{journal}{Phys. Rev.} \textbf{\bibinfo{volume}{176}},
  \bibinfo{pages}{722} (\bibinfo{year}{1968}),
  \urlprefix\url{https://link.aps.org/doi/10.1103/PhysRev.176.722}.

\bibitem[{\citenamefont{Bernhoeft et~al.}(1998)\citenamefont{Bernhoeft, Hiess,
  Langridge, Stunault, Wermeille, Vettier, Lander, Huth, Jourdan, and
  Adrian}}]{Bernhoeft1998}
\bibinfo{author}{\bibfnamefont{N.}~\bibnamefont{Bernhoeft}},
  \bibinfo{author}{\bibfnamefont{A.}~\bibnamefont{Hiess}},
  \bibinfo{author}{\bibfnamefont{S.}~\bibnamefont{Langridge}},
  \bibinfo{author}{\bibfnamefont{A.}~\bibnamefont{Stunault}},
  \bibinfo{author}{\bibfnamefont{D.}~\bibnamefont{Wermeille}},
  \bibinfo{author}{\bibfnamefont{C.}~\bibnamefont{Vettier}},
  \bibinfo{author}{\bibfnamefont{G.~H.} \bibnamefont{Lander}},
  \bibinfo{author}{\bibfnamefont{M.}~\bibnamefont{Huth}},
  \bibinfo{author}{\bibfnamefont{M.}~\bibnamefont{Jourdan}}, \bibnamefont{and}
  \bibinfo{author}{\bibfnamefont{H.}~\bibnamefont{Adrian}},
  \bibinfo{journal}{Physical Review Letters} \textbf{\bibinfo{volume}{81}},
  \bibinfo{pages}{3419} (\bibinfo{year}{1998}), ISSN \bibinfo{issn}{10797114}.

\bibitem[{\citenamefont{Masaki et~al.}(1972)\citenamefont{Masaki, Tagawa, and
  Tsuji}}]{Masaki1972}
\bibinfo{author}{\bibfnamefont{N.}~\bibnamefont{Masaki}},
  \bibinfo{author}{\bibfnamefont{H.}~\bibnamefont{Tagawa}}, \bibnamefont{and}
  \bibinfo{author}{\bibfnamefont{T.}~\bibnamefont{Tsuji}},
  \bibinfo{journal}{Journal of Nuclear Materials}
  \textbf{\bibinfo{volume}{45}}, \bibinfo{pages}{230} (\bibinfo{year}{1972}).

\bibitem[{\citenamefont{Tobisch and Hase}(1967)}]{Tobisch1967}
\bibinfo{author}{\bibfnamefont{J.}~\bibnamefont{Tobisch}} \bibnamefont{and}
  \bibinfo{author}{\bibfnamefont{W.}~\bibnamefont{Hase}},
  \bibinfo{journal}{physica status solidi (b)} \textbf{\bibinfo{volume}{21}}
  (\bibinfo{year}{1967}).

\bibitem[{\citenamefont{Collins et~al.}(2001)\citenamefont{Collins, Laundy, and
  Stunault}}]{Collins2001}
\bibinfo{author}{\bibfnamefont{S.~P.} \bibnamefont{Collins}},
  \bibinfo{author}{\bibfnamefont{D.}~\bibnamefont{Laundy}}, \bibnamefont{and}
  \bibinfo{author}{\bibfnamefont{A.}~\bibnamefont{Stunault}},
  \bibinfo{journal}{J. Phys.: Condens. Matter} \textbf{\bibinfo{volume}{13}},
  \bibinfo{pages}{1891} (\bibinfo{year}{2001}).

\bibitem[{\citenamefont{Kokubun and Dmitrienko}(2012)}]{Kokubun2012}
\bibinfo{author}{\bibfnamefont{J.}~\bibnamefont{Kokubun}} \bibnamefont{and}
  \bibinfo{author}{\bibfnamefont{V.~E.} \bibnamefont{Dmitrienko}},
  \bibinfo{journal}{European Physical Journal: Special Topics}
  \textbf{\bibinfo{volume}{208}}, \bibinfo{pages}{39} (\bibinfo{year}{2012}),
  ISSN \bibinfo{issn}{19516355}.

\bibitem[{\citenamefont{Santini et~al.}(2009)\citenamefont{Santini, Carretta,
  Amoretti, Caciuffo, Magnani, and Lander}}]{Santini2009}
\bibinfo{author}{\bibfnamefont{P.}~\bibnamefont{Santini}},
  \bibinfo{author}{\bibfnamefont{S.}~\bibnamefont{Carretta}},
  \bibinfo{author}{\bibfnamefont{G.}~\bibnamefont{Amoretti}},
  \bibinfo{author}{\bibfnamefont{R.}~\bibnamefont{Caciuffo}},
  \bibinfo{author}{\bibfnamefont{N.}~\bibnamefont{Magnani}}, \bibnamefont{and}
  \bibinfo{author}{\bibfnamefont{G.~H.} \bibnamefont{Lander}},
  \bibinfo{journal}{Reviews of Modern Physics} \textbf{\bibinfo{volume}{81}},
  \bibinfo{pages}{807} (\bibinfo{year}{2009}), ISSN \bibinfo{issn}{00346861}.

\bibitem[{\citenamefont{Wedgwood}(1974)}]{Wedgwood1974}
\bibinfo{author}{\bibfnamefont{F.~A.} \bibnamefont{Wedgwood}},
  \bibinfo{journal}{Journal of Physics C: Solid State Physics}
  \textbf{\bibinfo{volume}{7}}, \bibinfo{pages}{3203} (\bibinfo{year}{1974}).

\bibitem[{\citenamefont{Aczel et~al.}(2012)\citenamefont{Aczel, Granroth,
  Macdougall, Buyers, Abernathy, Samolyuk, Stocks, and Nagler}}]{Aczel2012a}
\bibinfo{author}{\bibfnamefont{A.~A.} \bibnamefont{Aczel}},
  \bibinfo{author}{\bibfnamefont{G.~E.} \bibnamefont{Granroth}},
  \bibinfo{author}{\bibfnamefont{G.~J.} \bibnamefont{Macdougall}},
  \bibinfo{author}{\bibfnamefont{W.~J.~L.} \bibnamefont{Buyers}},
  \bibinfo{author}{\bibfnamefont{D.~L.} \bibnamefont{Abernathy}},
  \bibinfo{author}{\bibfnamefont{G.~D.} \bibnamefont{Samolyuk}},
  \bibinfo{author}{\bibfnamefont{G.~M.} \bibnamefont{Stocks}},
  \bibnamefont{and} \bibinfo{author}{\bibfnamefont{S.~E.}
  \bibnamefont{Nagler}}, \bibinfo{journal}{Nature communications}
  \textbf{\bibinfo{volume}{3}}, \bibinfo{pages}{1124} (\bibinfo{year}{2012}),
  ISSN \bibinfo{issn}{2041-1723},
  \urlprefix\url{http://www.ncbi.nlm.nih.gov/pubmed/23047682}.

\bibitem[{\citenamefont{Lin et~al.}(2014)\citenamefont{Lin, Aczel, Abernathy,
  Nagler, Buyers, and Granroth}}]{Lin2014}
\bibinfo{author}{\bibfnamefont{J.~Y.} \bibnamefont{Lin}},
  \bibinfo{author}{\bibfnamefont{A.~A.} \bibnamefont{Aczel}},
  \bibinfo{author}{\bibfnamefont{D.~L.} \bibnamefont{Abernathy}},
  \bibinfo{author}{\bibfnamefont{S.~E.} \bibnamefont{Nagler}},
  \bibinfo{author}{\bibfnamefont{W.~J.} \bibnamefont{Buyers}},
  \bibnamefont{and} \bibinfo{author}{\bibfnamefont{G.~E.}
  \bibnamefont{Granroth}}, \bibinfo{journal}{Physical Review B}
  \textbf{\bibinfo{volume}{89}}, \bibinfo{pages}{144302}
  (\bibinfo{year}{2014}), ISSN \bibinfo{issn}{1550235X}.

\bibitem[{\citenamefont{Brooks and Kelly}(1983)}]{Brooks1983}
\bibinfo{author}{\bibfnamefont{M.~S.} \bibnamefont{Brooks}} \bibnamefont{and}
  \bibinfo{author}{\bibfnamefont{P.~J.} \bibnamefont{Kelly}},
  \bibinfo{journal}{Physical Review Letters} \textbf{\bibinfo{volume}{51}},
  \bibinfo{pages}{1708} (\bibinfo{year}{1983}), ISSN \bibinfo{issn}{00319007}.

\bibitem[{\citenamefont{Holden et~al.}(1984)\citenamefont{Holden, Buyers,
  Svensson, and Lander}}]{Holden1984}
\bibinfo{author}{\bibfnamefont{T.~M.} \bibnamefont{Holden}},
  \bibinfo{author}{\bibfnamefont{W.~J.~L.} \bibnamefont{Buyers}},
  \bibinfo{author}{\bibfnamefont{E.~C.} \bibnamefont{Svensson}},
  \bibnamefont{and} \bibinfo{author}{\bibfnamefont{G.~H.}
  \bibnamefont{Lander}}, \bibinfo{journal}{Physical Review B}
  \textbf{\bibinfo{volume}{30}}, \bibinfo{pages}{114} (\bibinfo{year}{1984}).

\bibitem[{\citenamefont{Fujimori et~al.}(2012)\citenamefont{Fujimori, Ohkochi,
  Okane, Saitoh, Fujimori, and Yamagami}}]{Fujimori2012}
\bibinfo{author}{\bibfnamefont{S.-i.} \bibnamefont{Fujimori}},
  \bibinfo{author}{\bibfnamefont{T.}~\bibnamefont{Ohkochi}},
  \bibinfo{author}{\bibfnamefont{T.}~\bibnamefont{Okane}},
  \bibinfo{author}{\bibfnamefont{Y.}~\bibnamefont{Saitoh}},
  \bibinfo{author}{\bibfnamefont{A.}~\bibnamefont{Fujimori}}, \bibnamefont{and}
  \bibinfo{author}{\bibfnamefont{H.}~\bibnamefont{Yamagami}},
  \bibinfo{journal}{Phys. Rev. B} \textbf{\bibinfo{volume}{86}},
  \bibinfo{pages}{235108} (\bibinfo{year}{2012}).

\bibitem[{\citenamefont{Fujimori et~al.}(2016)\citenamefont{Fujimori, Takeda,
  Okane, Saitoh, Fujimori, Yamagami, Haga, Yamamoto, and Onuki}}]{Fujimori2016}
\bibinfo{author}{\bibfnamefont{S.}~\bibnamefont{Fujimori}},
  \bibinfo{author}{\bibfnamefont{Y.}~\bibnamefont{Takeda}},
  \bibinfo{author}{\bibfnamefont{T.}~\bibnamefont{Okane}},
  \bibinfo{author}{\bibfnamefont{Y.}~\bibnamefont{Saitoh}},
  \bibinfo{author}{\bibfnamefont{A.}~\bibnamefont{Fujimori}},
  \bibinfo{author}{\bibfnamefont{H.}~\bibnamefont{Yamagami}},
  \bibinfo{author}{\bibfnamefont{Y.}~\bibnamefont{Haga}},
  \bibinfo{author}{\bibfnamefont{E.}~\bibnamefont{Yamamoto}}, \bibnamefont{and}
  \bibinfo{author}{\bibfnamefont{Y.}~\bibnamefont{Onuki}}, \bibinfo{journal}{J.
  Phys. Soc. Jpn.} \textbf{\bibinfo{volume}{85}} (\bibinfo{year}{2016}).

\bibitem[{\citenamefont{Gorbunov et~al.}(2019)\citenamefont{Gorbunov, Nomura,
  Zvyagin, Henriques, Andreev, Skourski, Zvyagina, Tro{\'{c}}, Zherlitsyn, and
  Wosnitza}}]{Gorbunov2019}
\bibinfo{author}{\bibfnamefont{D.~I.} \bibnamefont{Gorbunov}},
  \bibinfo{author}{\bibfnamefont{T.}~\bibnamefont{Nomura}},
  \bibinfo{author}{\bibfnamefont{A.~A.} \bibnamefont{Zvyagin}},
  \bibinfo{author}{\bibfnamefont{M.~S.} \bibnamefont{Henriques}},
  \bibinfo{author}{\bibfnamefont{A.~V.} \bibnamefont{Andreev}},
  \bibinfo{author}{\bibfnamefont{Y.}~\bibnamefont{Skourski}},
  \bibinfo{author}{\bibfnamefont{G.~A.} \bibnamefont{Zvyagina}},
  \bibinfo{author}{\bibfnamefont{R.}~\bibnamefont{Tro{\'{c}}}},
  \bibinfo{author}{\bibfnamefont{S.}~\bibnamefont{Zherlitsyn}},
  \bibnamefont{and} \bibinfo{author}{\bibfnamefont{J.}~\bibnamefont{Wosnitza}},
  \bibinfo{journal}{Phys. Rev. B} \textbf{\bibinfo{volume}{100}},
  \bibinfo{pages}{024417} (\bibinfo{year}{2019}).

\end{thebibliography}

\end{document}